\documentclass{aa}

\newcommand\dosingle[1]{#1}  \newcommand\dodouble[1]{ } 

\newcommand\nice[1]{#1}    \newcommand\subm[1]{}   
\subm{ \renewcommand\dosingle[1]{ }   \renewcommand\dodouble[1]{#1} }

\dodouble{ \documentclass[referee]{aa} }

\usepackage{graphicx}

\usepackage{amsfonts}
\usepackage{mathpi}

\usepackage{url} 

\newcommand\postrefereechanges[1]{#1}

 \usepackage{hyperref}
\nice{ 
\providecommand{\eprint}[1]{\href{http://arxiv.org/abs/#1}{{\tt [arXiv:#1]}}}

\providecommand{\url}[1]{\href{#1}{#1}}

}

\providecommand{\adsurl}[1]{} 
\newcommand\SSS{Sect.~}

\nice{ \newcommand\mycaptionfont{\protect\footnotesize} }

\newcommand\kms{\,km\,s$^{-1}$}

\newcommand\gtapprox{\,\lower.6ex\hbox{$\buildrel >\over \sim$} \, }
\newcommand\ltapprox{\,\lower.6ex\hbox{$\buildrel <\over \sim$} \, }
\newcommand\propapprox{\,\lower.6ex\hbox{$\buildrel \propto\over \sim$} \, }

\newcommand\arcs{\ifmmode {'' }\else $'' $\fi}     
\newcommand\arcm{\ifmmode {' }\else $' $\fi}       
\newcommand\ddeg{\ifmmode^\circ\else$^\circ$\fi}    

\newcommand\frtoday{Le\space\number\day\space\ifcase\month\or
  janvier\or f\'evrier\or mars\or avril\or mai\or juin\or
  juillet\or ao\^ut\or septembre\or octobre\or novembre\or 
d\'ecembre\fi\space \number\year}

\newcommand\cqg{ClassQuantGra}   %
\newcommand\hGpc{\mbox{$h^{-1}$ Gpc}}

\newcommand\rSLS{r_{\mathrm{SLS}}}  
\newcommand\rC{R_{\mathrm{C}}}  
\newcommand\rSLStiny{r_{\mathrm{SLS}}}  

\newcommand\Omm{\Omega_{\mathrm{m}}}
\newcommand\Omtot{\Omega_{\mathrm{tot}}}

\newcommand\ximc{\xi_{\mathrm{C}}} 
\newcommand\xisc{\xi_{\mathrm{A}}} 

\newcommand\Npoint{N_{\mathrm{p}}}

\newcommand\Nmc{N_{\mathrm{C}}} 
\newcommand\Nsc{N_{\mathrm{A}}} 

\newcommand\Nchain{N_{\mathrm{chain}}}

\newcommand\Pmin{P_{\mathrm{min}}}

\newcommand\Sxi{S_\xi}  
\newcommand\SxiINCthr{S_\xi^{\mathrm{INC3}}}  
\newcommand\SxiWMAP{S_\xi^{\mathrm{WMAP}}}  
\newcommand\alphaINCthr{\alpha^{\mathrm{INC3}}}  
\newcommand\alphalimit{\alpha^{\mathrm{limit}}}  
\newcommand\phiINCthr{\phi^{\mathrm{INC3}}}  
\newcommand\phirms{\phi_{\mathrm{r.m.s.}}}  

\newcommand\lII{l}
\newcommand\bII{b}

\newcommand\sigmalbth{\sigma_{\left<(l,b)\right>}}

\newcommand\PDSlikesignal{\ximc^{\mathrm{WMAP}}}

\newcommand\notea{^\mathrm{a}}
\newcommand\noteb{^\mathrm{b}}
\newcommand\notec{^\mathrm{c}}
\newcommand\noted{^\mathrm{d}}

\title{Poincar\'e 
dodecahedral space parameter estimates}

\author{Boudewijn F. Roukema\inst{1} \and
Zbigniew Buli\'nski\inst{1} \and
Nicolas E. Gaudin\inst{2}  
}

\institute{Toru\'n Centre for Astronomy, Nicolaus Copernicus University,
ul. Gagarina 11, 87-100 Toru\'n, Poland 
\and
{\'Ecole nationale sup\'erieure de physique de
Strasbourg, Universit\'e Louis Pasteur, 
Bd. S\'ebastien Brant, BP 10413, 67412 Illkirch Cedex,
France}
}

\date{\frtoday}

\titlerunning{Poincar\'e dodecahedral space parameters}
\authorrunning{Roukema, Buli\'nski \& Gaudin}

\begin{document}


\newcommand\Nchainsmain{16}
\newcommand\Npergroup{four}

\abstract
{
Several studies have proposed that the preferred
model of the comoving spatial 3-hypersurface of the Universe 
may be a Poincar\'e
dodecahedral space (PDS) rather than
a simply connected, infinite, flat space. 
}
{
Here, we aim to improve the
surface of last scattering (SLS) optimal cross-correlation method
and apply this to observational data and simulations.
}
{
For a given ``generalised'' 
PDS orientation, we analytically derive the formulae required to 
exclude points on the sky that cannot be members of close SLS-SLS 
cross-pairs. These enable more efficient pair selection without
sacrificing the uniformity of the underlying selection process. For 
a sufficiently small matched circle size $\alpha$ and 
a fixed number of randomly placed points selected for a cross-correlation
estimate, the calculation time is decreased and the 
number of pairs per separation bin is increased.
Using this faster method, and including
the smallest separation bin when testing correlations, 
(i) we recalculate Monte Carlo Markov Chains (MCMC) on the five-year
Wilkinson Microwave Anisotropy Probe (WMAP) data; and
(ii) we seek PDS solutions in a small number of
Gaussian random fluctuation (GRF) simulations
in order 
to further explore the statistical significance of the PDS hypothesis.
}
{
For $5\ddeg < \alpha < 60\ddeg$, a calculation speed-up of 3--10 is obtained.
(i) The best estimates of the PDS parameters for the five-year WMAP 
data are similar to those for the three-year data. 
(ii) Comparison of the optimal solutions found by the 
MCMC chains in the observational map to those found in the simulated maps
yields a slightly stronger rejection of the simply connected model using $\alpha$
rather than the twist angle $\phi$.
The best estimate of $\alpha$ implies that,
{{\em given a large-scale auto-correlation as weak as that observed,}}
the PDS-like cross-correlation 
signal in the WMAP data is expected with a probability of less than 
about 10\%. The expected distribution of $\phi$ from the GRF simulations
is not uniform on $[-\pi,\pi]$. 
%
}
{Using this faster algorithm, we find that the previous PDS parameter
  estimates are stable to the update to five-year WMAP data.
  Moreover, for an infinite, flat, cosmic concordance model with
  Gaussian random fluctuations, the chance of finding {{\em both}} (a)
  a large-scale auto-correlation as weak as observed and (b) a
  PDS-like signal similar to what is observed is less than about 0.015\%
  to 1.25\%.  }


\maketitle

\dodouble{ \clearpage } 


\newcommand\tbench{
\begin{table}
\caption{\mycaptionfont 
Example of benchmarking on a 3GHz processor.$\notea$
%
%
\label{t-bench}}
$$\begin{array}{r c r r r r r} \hline   \hline 
\alpha &
r_2 &
\alpha_+\noteb &
\Npoint \rule[-1.5ex]{0ex}{4.5ex} &
 t\notec & \Nsc\noted & \Nmc\noted \\
{}\ddeg &\approx {\hGpc} & {}\ddeg &  &  \mathrm{s} & & \\
\hline \rule{0ex}{2.5ex}  
 5 &  0.4 & \ldots &  1000 &          5 &     1317 & 52 \\
 5 &  0.4 & \ldots &  2000 &           22 &    5268 & 206 \\
 5 &  0.4 & \ldots  & 8000 &           375 &   83476 & 3900 \\
%
 5 & 4.4 & \ldots &  1000 &               5 &     1317 & 52 \\
 5 & 4.4 & \ldots &  2000 &               23 &    5268 & 206 \\
 5 & 4.4 & \ldots &  8000 &               377 &   83476 & 3900 \\
\hline \rule{0ex}{2.5ex}  
 5 & 0.4 &  \;  13.6 & 1000 &        0 &     6994 & 1911 \\
  5 & 0.4 &  13.6  & 2000   &       2 &     28085 & 7860 \\
 5 & 0.4 &  13.6  & 8000   &       41 &    449308 & 126395 \\
 5 & 4.4 &  45.0  & 1000   &           0 &     1317 & 52 \\
5 & 4.4 &  45.0  & 2000   &           3 &     5268 & 204 \\
 5 & 4.4 &  45.0  & 8000   &           53 &    83476 & 3880 \\
 \hline \rule{0ex}{2.5ex}  
 60 & 0.4 & \ldots &  1000 &           5 &     1408 & 192 \\
 60 & 0.4 & \ldots &  2000 &           23 &    5650 & 736 \\
 60 & 0.4 & \ldots &  8000 &           354 &   90242 & 12070 \\
%
 60 & 4.4 & \ldots &  1000 &            5 &     1408 & 192 \\
 60 & 4.4 & \ldots &  2000 &            24 &    5650 & 736 \\
 60 & 4.4 & \ldots &  8000 &            369 &   90242 & 12070 \\
\hline \rule{0ex}{2.5ex}  
 60 & 0.4 &  62.9  & 1000   &       0 &     3532 & 1518 \\
 60 & 0.4 &  62.9  & 2000   &       3 &     14290 & 5986 \\
 60 & 0.4 &  62.9  & 8000   &       48 &    228426 & 94887 \\
 60 & 4.4 &  90.9  & 1000   &         2 &     1408 & 183 \\
 60 & 4.4 &  90.9  & 2000   &         8 &     5650 & 706 \\
 60 & 4.4 &  90.9  & 8000   &         130 &   90242 & 11536 \\
\hline
\end{array}$$
\protect\postrefereechanges{ 
\\ 
$\notea$ The pseudo-random number generator
has the same initial seed for each calculation. \\
$\noteb$ An estimate of $\alpha_+$ from Eq.~(\protect\ref{e-alpha-pm}) 
is shown in the cases where 
pair preselection as described in 
\SSS\protect\ref{s-method-anglimits} is used.  \\
$\notec$ Calculation time. \\
$\noted$ Numbers of pairs in the smallest bin $\Nsc, \Nmc$ 
for $\xisc, \ximc$ respectively. \\
}
\end{table}
}  

\newcommand\tdodec{
\begin{table}
\caption{\mycaptionfont 
Sky positions of the best estimate of 
the six dodecahedral face centres for the five-year ILC map with the kp2 
mask.
%
\label{t-dodec}}
$$\begin{array}{c c r  r r r} \hline \hline
\rule[-1.5ex]{0ex}{4.5ex}
\Pmin &
i\notea & 
n & 
{\lII}  & {\bII} 
& \sigmalbth \\
 &
 & 
 & 
 {}\ddeg & {}\ddeg 
&  {}\ddeg\\
 \hline 
\rule{0ex}{2.5ex}
%
   0.4 &  7 &  5141 & \ \ 182.8 &   62.3 &    1.1  \\
   0.4 & 12 &  5239 &  305.5 &   44.4 &    1.2  \\
   0.4 &  3 &  7097 &   45.0 &   49.5 &    0.7  \\
   0.4 &  5 &  5812 &  115.9 &   19.5 &    1.2  \\
   0.4 &  8 &  5056 &  174.6 &   -3.0 &    2.9  \\
   0.4 & 10 &  4707 &  239.8 &   13.9 &    2.1  \\
\hline \rule{0ex}{2.5ex}
   0.5\noteb  &  7 &  2736 &  181.1 &   62.2 &    1.4  \\
   0.5 & 12 &  2838 &  305.7 &   44.5 &    1.3  \\
   0.5 &  3 &  3712 &   45.4 &   49.3 &    0.7  \\
   0.5 &  5 &  3005 &  114.6 &   18.7 &    1.6  \\
   0.5 &  8 &  2732 &  178.1 &   -0.5 &    1.6  \\
   0.5 & 10 &  2570 &  239.9 &   13.8 &    1.6  \\
\hline \rule{0ex}{2.5ex}
   0.6 &  7 &  1429 &  179.3 &   62.6 &    1.1  \\
   0.6 & 12 &  1487 &  306.0 &   44.3 &    1.2  \\
   0.6 &  3 &  1919 &   45.5 &   48.8 &    0.8  \\
   0.6 &  5 &  1544 &  113.6 &   18.4 &    2.0  \\
   0.6 &  8 &  1387 &  175.6 &   -1.6 &    3.0  \\
   0.6 & 10 &  1419 &  237.7 &   14.5 &    1.6  \\
\hline
\end{array}$$
\\ 
\postrefereechanges{
$\notea$ The face centres are ordered according to the ordering in 
Table~1 of RBSG08. The other 6 faces are directly opposite with identical 
errors. \\
$\noteb$ The estimate for $\Pmin = 0.5$ corresponds 
to the points shown in the upper panel of Fig.~\protect\ref{f-five_lbth_N}, based on 
160,000 steps of MCMC chains. \\
}
\end{table}
}  

\newcommand\talphaphifiveyear{
\begin{table}
\caption{\mycaptionfont 
Estimates of matched circle radius $\alpha$ and twist
phase $\phi$ from the Integrated Linear Combination (ILC) 
and \protect\nocite{WMAPTegmarkFor}{Tegmark} {et~al.} (2003) (TOH) versions of the 
five-year WMAP data.
\label{t-alpha-phi-5yr}}
$$\begin{array}{c c c r r r r} \hline \hline
\mbox{map}
&
\Pmin \rule[-1.5ex]{0ex}{4.5ex}
&
n^{\mathrm{a}} & 
\alpha & \sigma_{\left<\alpha\right>} &
 \phi  & \sigma_{\left<\phi\right>} 
\\ 
&
&
 & 
{}\ddeg & {}\ddeg & {}\ddeg & {}\ddeg  
\\
\hline 
\rule{0ex}{2.5ex}     
%
\mbox{ILC} & 0.4 & 5508.67 & 20.3 & 0.7 & 38.7 & 1.6 \\
\mbox{ILC} & 0.5 & 2932.17 & 20.5 & 1 & 37.8 & 1.2 \\
\mbox{ILC} & 0.6 & 1530.83 & 20.1 & 0.7 & 39.8 & 1.1 \\
\hline \rule{0ex}{2.5ex}     
%
\mbox{TOH} & 0.4 & 4802.08 & 21.2 & 1.4 & 30.6 & 4.8 \\
\mbox{TOH} & 0.5 & 2955.58 & 20.4 & 0.8 & 32.1 & 2.6 \\
\mbox{TOH} & 0.6 & 1655.92 & 20.3 & 0.7 & 27.5 & 5.5 \\
\hline
\end{array}$$
\begin{list}{}{}
\item[$^{\mathrm{a}}$] The number of steps $n$ can be a non-integer since for a given
MCMC step, it is possible that some of the face centres 
fall within the convergence radius of the final iteration 
as described in \SSS\protect\ref{s-res-lbtheta}, but other face centres do not.
\end{list}
\end{table}
}  

\newcommand\talphaphiINCthr{
\begin{table}
\caption{\mycaptionfont 
Estimates of matched circle radius $\alpha$ and twist
phase $\phi$ for the INC3 observational map. 
\label{t-alpha-phi-INC3}}
$$\begin{array}{c r r r} \hline  \hline
\Pmin\notea \rule[-1.5ex]{0ex}{4.5ex}
&
n\noteb & 
\alpha  &
 \phi 
\\ 
&
 & 
{}\ddeg &
{}\ddeg 
\\ \hline 
\rule{0ex}{2.5ex}    
 0.3 & 294.3 & \ \ 32.7 & \ 27.6 \\
 0.4 & 197.2 & 30.8 & 31.2 \\
 0.5 & 108.8 & 30.8 & 32.5 \\
\hline
\end{array}$$
\\
\postrefereechanges{
$\notea$ Minimum probability. \\
$\noteb$ Number of MCMC steps contributing to the estimate. \\
}
\end{table}
}  

\newcommand\tphisims{
\begin{table}
\caption{\mycaptionfont 
\protect\postrefereechanges{Properties of the distribution of best estimates of $\phi$ from 
the 20 simulations.}
\label{t-phi-sims}}
$$\begin{array}{c r r } \hline  \hline
\Pmin \rule[-1.5ex]{0ex}{4.5ex} 
& \phirms\notea
& P_{\mathrm{KS}}\noteb \\
& {}\ddeg & 
\\ \hline 
\rule{0ex}{2.5ex}  
0.3 & 32.9 & 5.9\%
\\
0.4 & 38.5 & 13.1\%
 \\
0.5 & 38.4 & 32.8\%
\\
\hline
\end{array}$$
\\
\postrefereechanges{
$\notea$ Root mean square width of the distribution. \\
$\noteb$ Two-sided Kolmogorov-Smirnov probability
that the simulational values are consistent
with a Gaussian distribution centred on zero, of width $\sigma_\phi = \phirms$. \\
}
\end{table}
}  

\newcommand\tSxifiveyr{ 
\begin{table}
\caption{\mycaptionfont 
$\Sxi$ estimates with and without the kp2 mask for the ILC and
TOH five-year maps.
\label{t-Sxi-fiveyr}}
$$\begin{array}{c r r } \hline  \hline
\rule[-1.5ex]{0ex}{4.5ex} 
\mbox{map} & \mbox{kp2} & \mbox{no mask} 
\\
& (\mu K)^4 & (\mu K)^4 \\ 
\\ \hline 
\rule{0ex}{2.5ex}  
\mathrm{ILC} & 1012 & 4851 
\\
\mathrm{TOH} & 1136 & 2749 
\\
\hline
\end{array}$$
\end{table}
}  

\newcommand\falphap{
\begin{figure}
\centering 
\includegraphics[width=6cm]{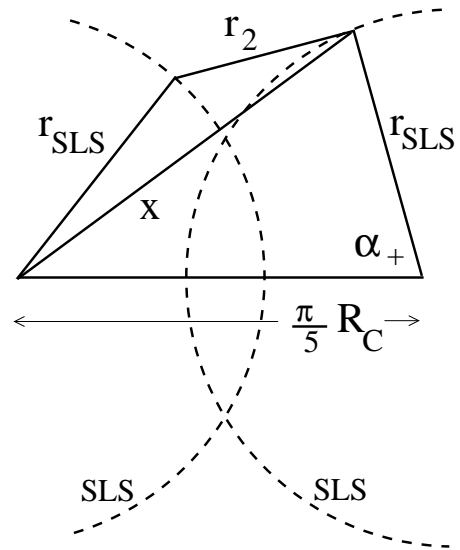}
\caption[]{ \mycaptionfont
Relation of a spatial geodesic of length $r_2$ joining a ``close''
pair of points in space to two copies of the SLS, and the angle
$\alpha_+$ separating the member of the pair on the right-hand SLS
from the dodecahedral face centre, i.e. from the spatial geodesic
joining the two SLS copies.  This figure
shows a spatial geodesic ``external'' to the matched circles (intersection
between the two copies of the SLS). 
The centres of the two copies of the SLS
(2-spheres) are separated by $(\pi/5) \rC$. 
See \SSS\protect\ref{s-method-anglimits}.
}
\label{f-alpha_p}
\end{figure} 
} 

\newcommand\falpham{
\begin{figure}
\centering 
\includegraphics[width=6cm]{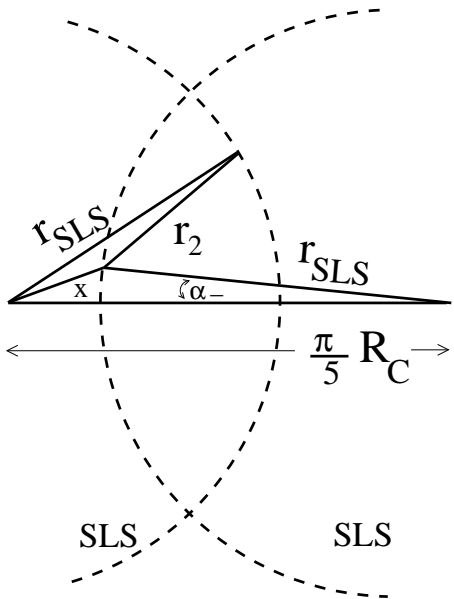}
\caption[]{ \mycaptionfont
As for Fig.~\protect\ref{f-alpha_p}, showing a spatial geodesic,
of length $r_2$,
``internal'' to the matched circles,  and angle
$\alpha_-$ separating the member of the pair on the right-hand SLS
from the dodecahedral face centre.
}
\label{f-alpha_m}
\end{figure} 
} 

\newcommand\falphaphicfsims{
\begin{figure}
\centering 
\includegraphics[width=8cm]{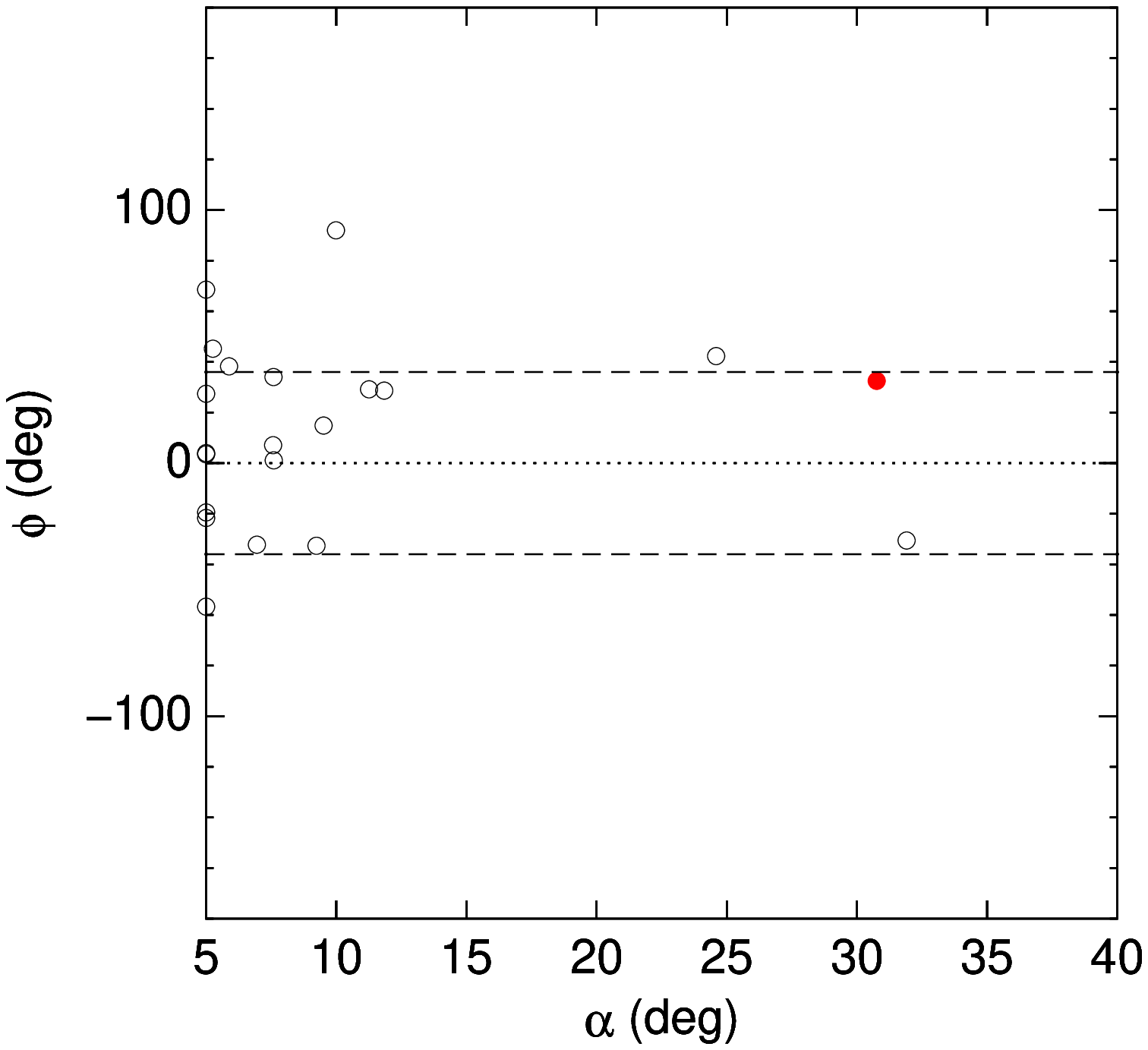}
\caption[]{ \mycaptionfont
Median circle sizes $\alpha$ 
(see \SSS\protect\ref{s-res-sim-alpha})
and twist angles $\phi$ (mean) for each of 
20 simulations (empty circles)
and the WMAP observational map (solid circle, 
values given in Table~\protect\ref{t-alpha-phi-INC3}),
\postrefereechanges{analysed using 
the steps with $P > 0.5$ [see Eqs~(25), (26) of 
RBSG08].} 
In each case, the 10,000 steps following 2000 burn-in steps 
of each of 4 MCMC chains started at random points in parameter
space are concatenated for this analysis.
}
\label{f-alpha_phi_cf_sims}
\end{figure} 
} 

\newcommand\fphihistsims{
\begin{figure}   
\centering 
\includegraphics[width=6cm]{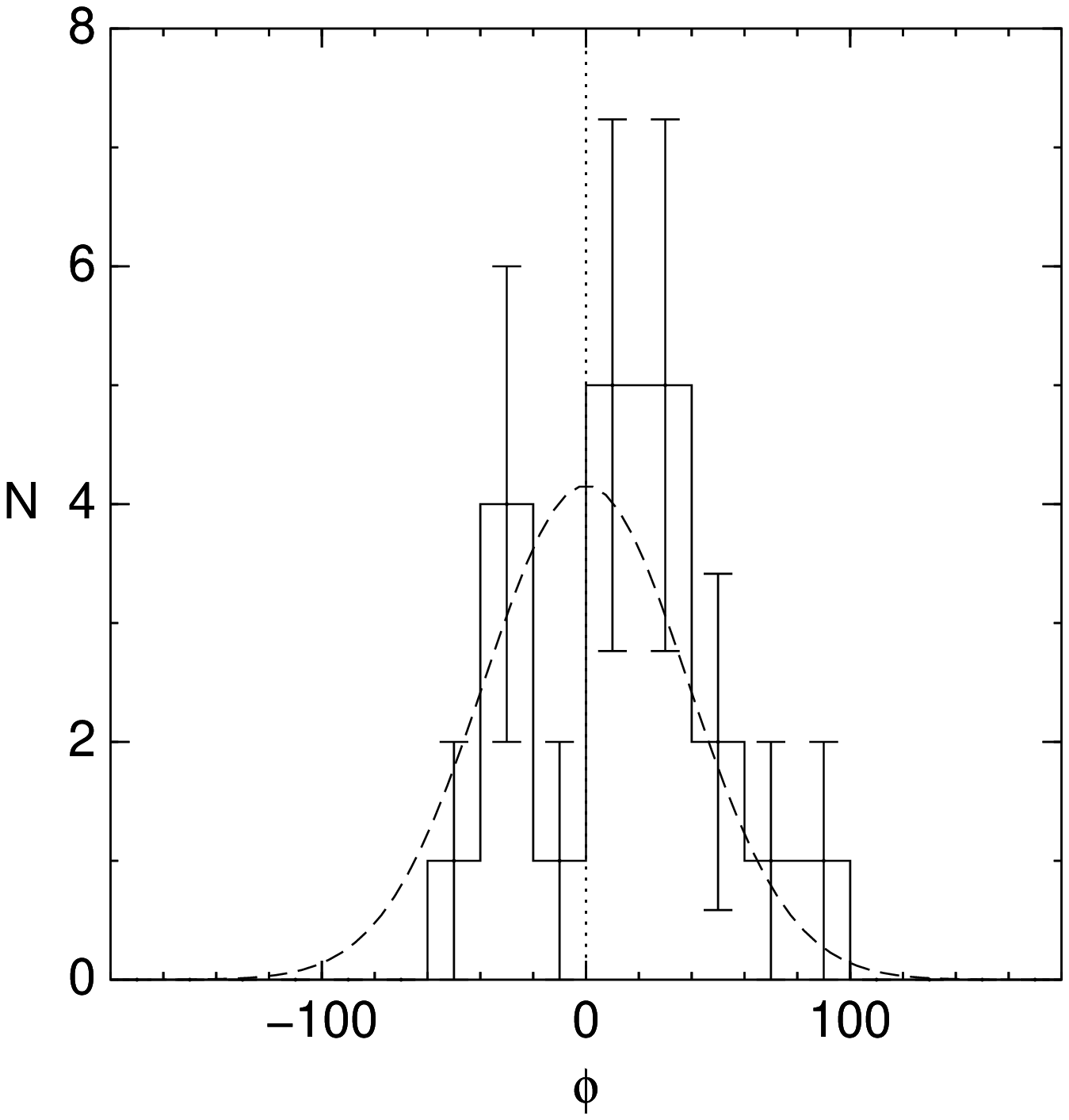}
\caption[]{ \mycaptionfont
\protect\postrefereechanges{Histogram of the optimal twist angle $\phi$ 
(shown in Fig.~\protect\ref{f-alpha_phi_cf_sims}), 
with $\sqrt{N}$ error bars, together
with a Gaussian distribution of width $38.4\ddeg$,
centred at zero
(see Table~\protect\ref{t-phi-sims}).}
}
\label{f-phi_hist_sims}
\end{figure} 
}  

\newcommand\falphaSsims{
\begin{figure}
\centering 
\includegraphics[width=8cm]{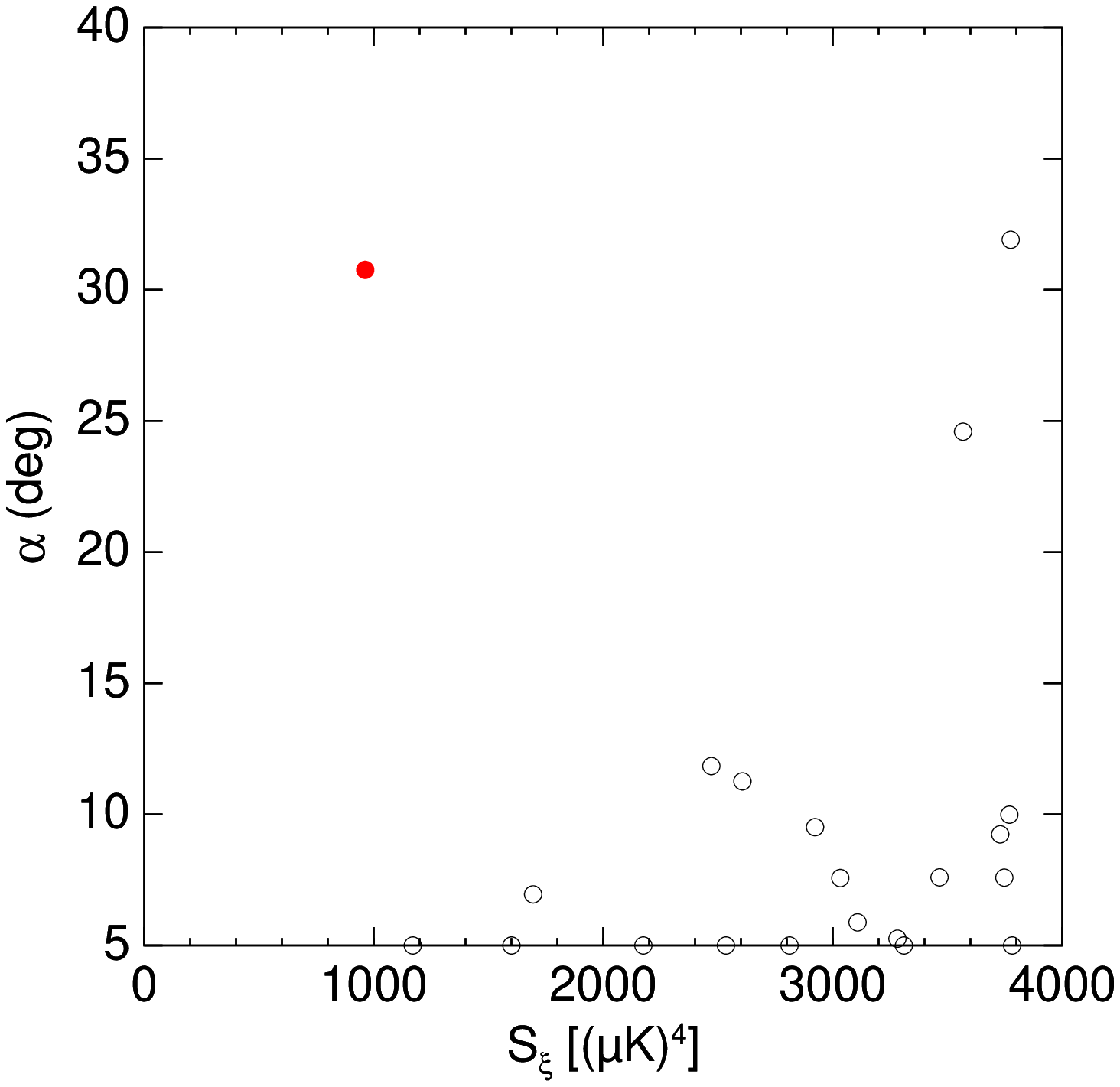}
\caption[]{ \mycaptionfont
Median circle sizes $\alpha$ as a function of 
$\Sxi$ for 
each of the
20 simulations (empty circles)
and the WMAP observational map (solid circle),
\postrefereechanges{analysed using the steps with $P > 0.5$.}
}
\label{f-alpha_S_sims}
\end{figure} 


} 

\newcommand\ffivelbthN{
\begin{figure}
\centering 
\includegraphics[width=6cm]{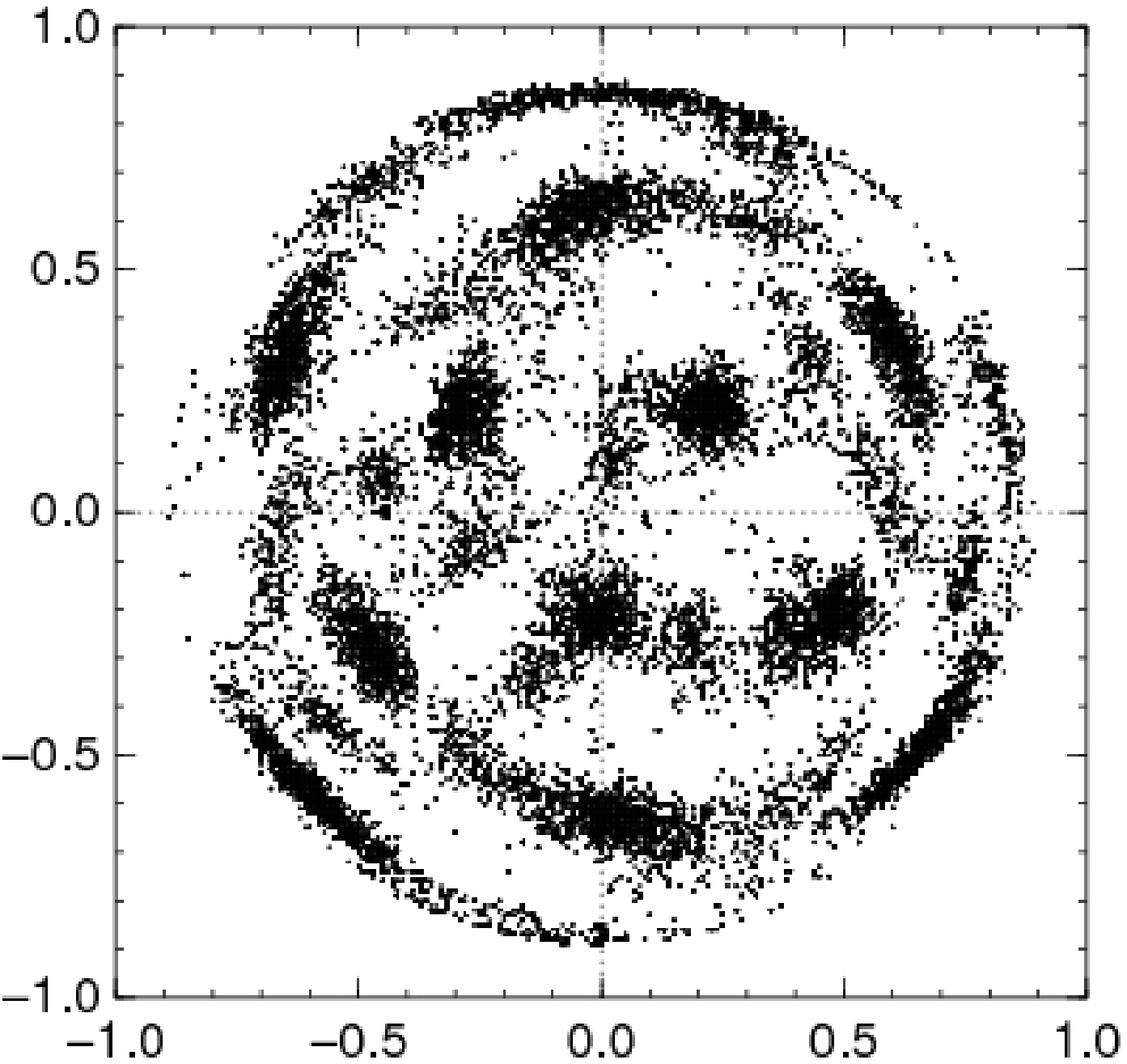} \\
\includegraphics[width=6cm]{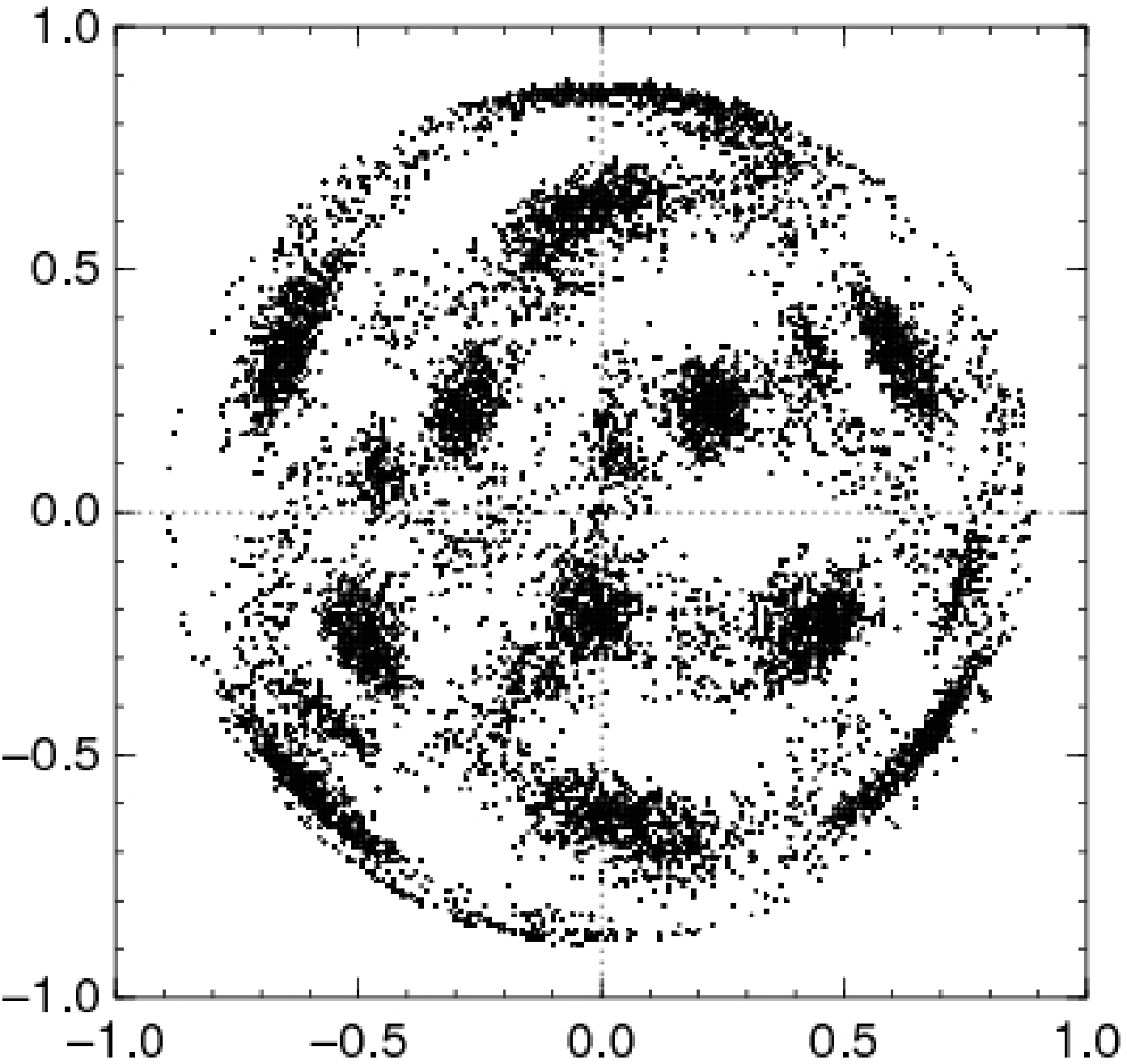}
\caption[]{ \mycaptionfont Full sky map 
  [Lambert azimuthal equal area projection
  \protect\nocite{Lambert1772}({Lambert} 1772), centred on the North
  Galactic Pole (NGP), with the $0\ddeg$ meridian as the positive vertical axis
  and galactic longitude increasing clockwise]
  showing the optimal
  orientation of dodecahedral face centres based on 
  {\Nchainsmain}0,000 steps in
  {\Nchainsmain} MCMC chains, using 
  \postrefereechanges{the five-year ILC map (upper panel) and 
  the five-year TOH map (lower panel),}
  and the kp2 mask, 
  showing face centres
  for which $P > 0.5$. 
}
\label{f-five_lbth_N}
\end{figure} 
} 

\newcommand\fINCthreelbthN{
\begin{figure}
\centering 
\includegraphics[width=6cm]{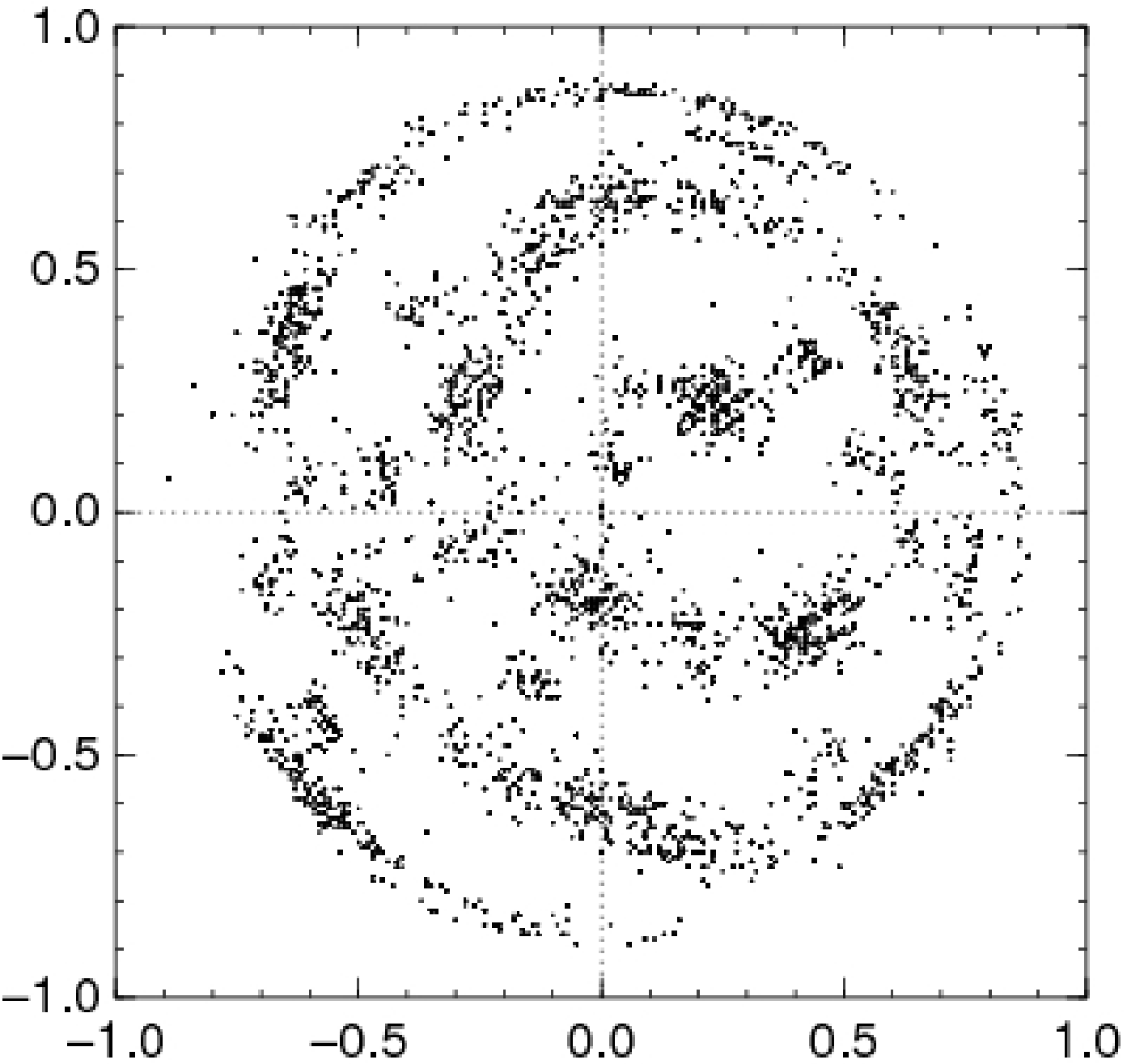}
\caption[]{ \mycaptionfont Full sky map 
  showing the optimal
  orientation of dodecahedral face centres based on 
  40,000 steps in
  4 MCMC chains, using the three-year INC map 
  and the kp2 mask, 
  showing face centres
  for which $P > 0.5$.
}
\label{f-INCthree_lbth_N}
\end{figure} 
} 

\newcommand\fINCthreelbthNsim{
\begin{figure}
\centering 
\includegraphics[width=6cm]{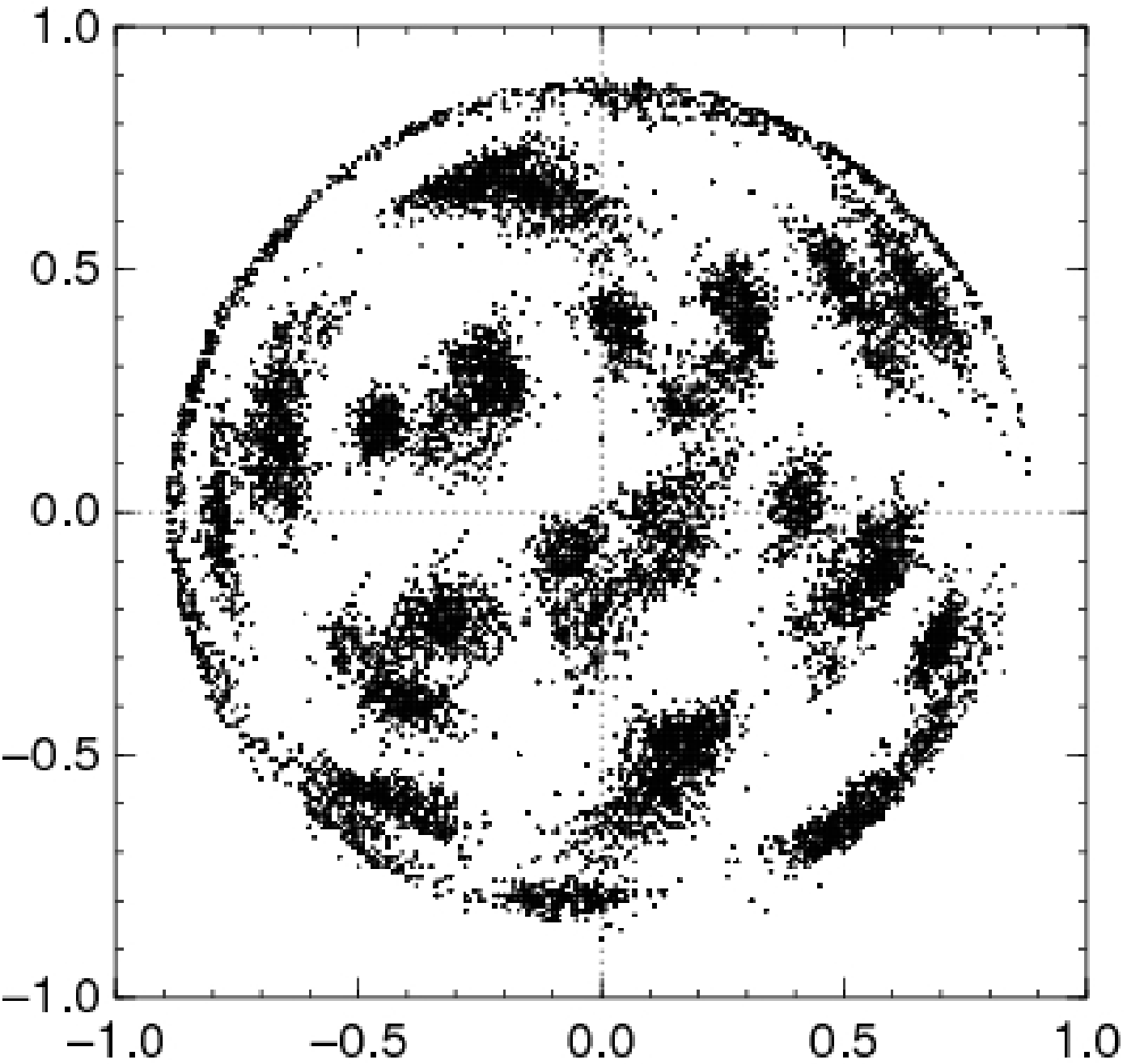} \\
\includegraphics[width=6cm]{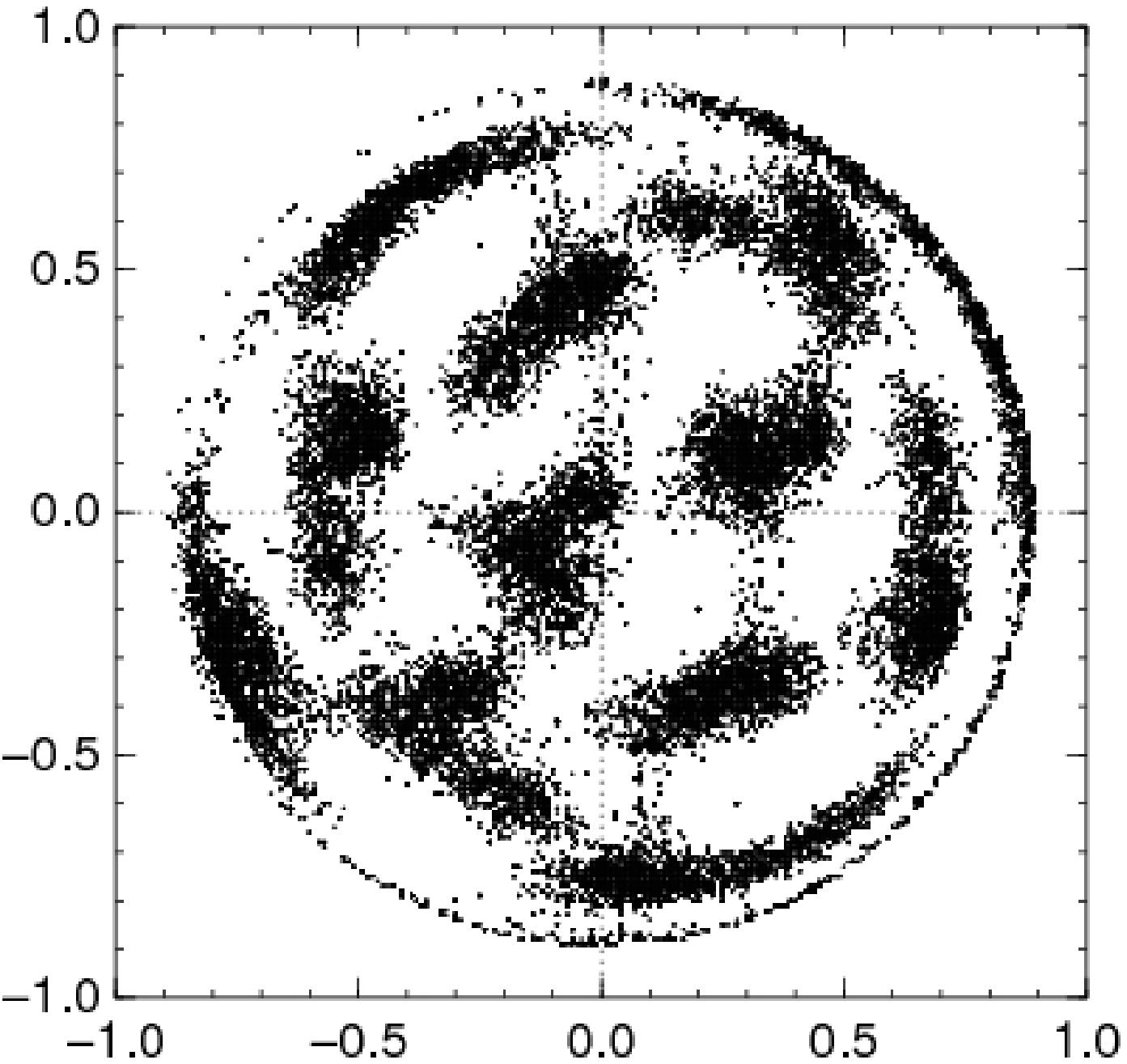}
\caption[]{ \mycaptionfont Full sky map 
  showing the optimal
  orientation of dodecahedral face centres based on 
  40,000 steps in
  4 MCMC chains, using the two simulated three-year INC maps
  with lowest $\Sxi$ estimates (simulations 92 and 90, in the upper and
  lower panels respectively),
  and the kp2 mask, 
  showing face centres
  for which $P > 0.5$.
}
\label{f-INCthree_lbth_N_sim}
\end{figure} 
} 

\newcommand\fINCthreelbthNsimhiS{
\begin{figure}
\centering 
\includegraphics[width=6cm]{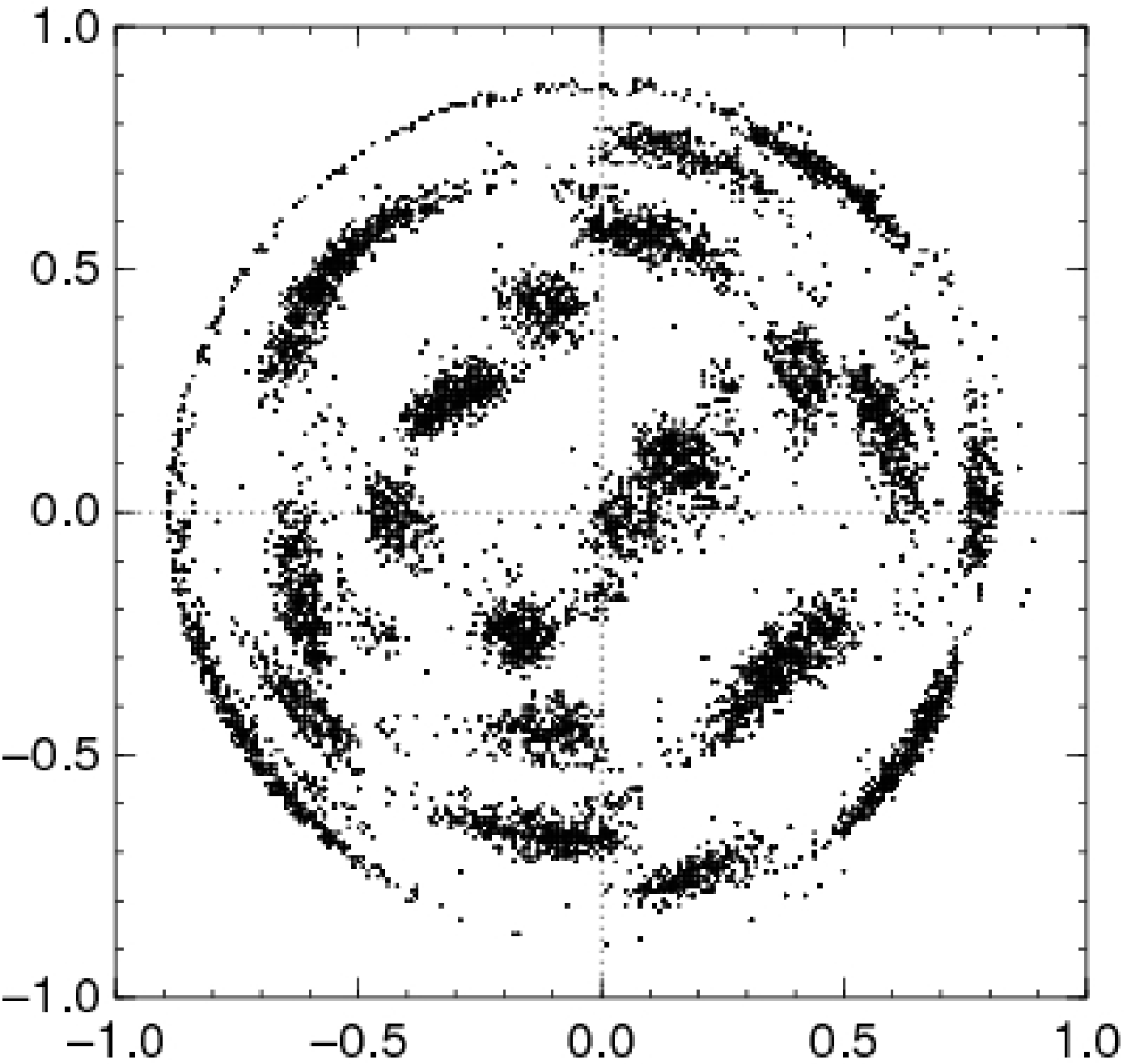} \\
\includegraphics[width=6cm]{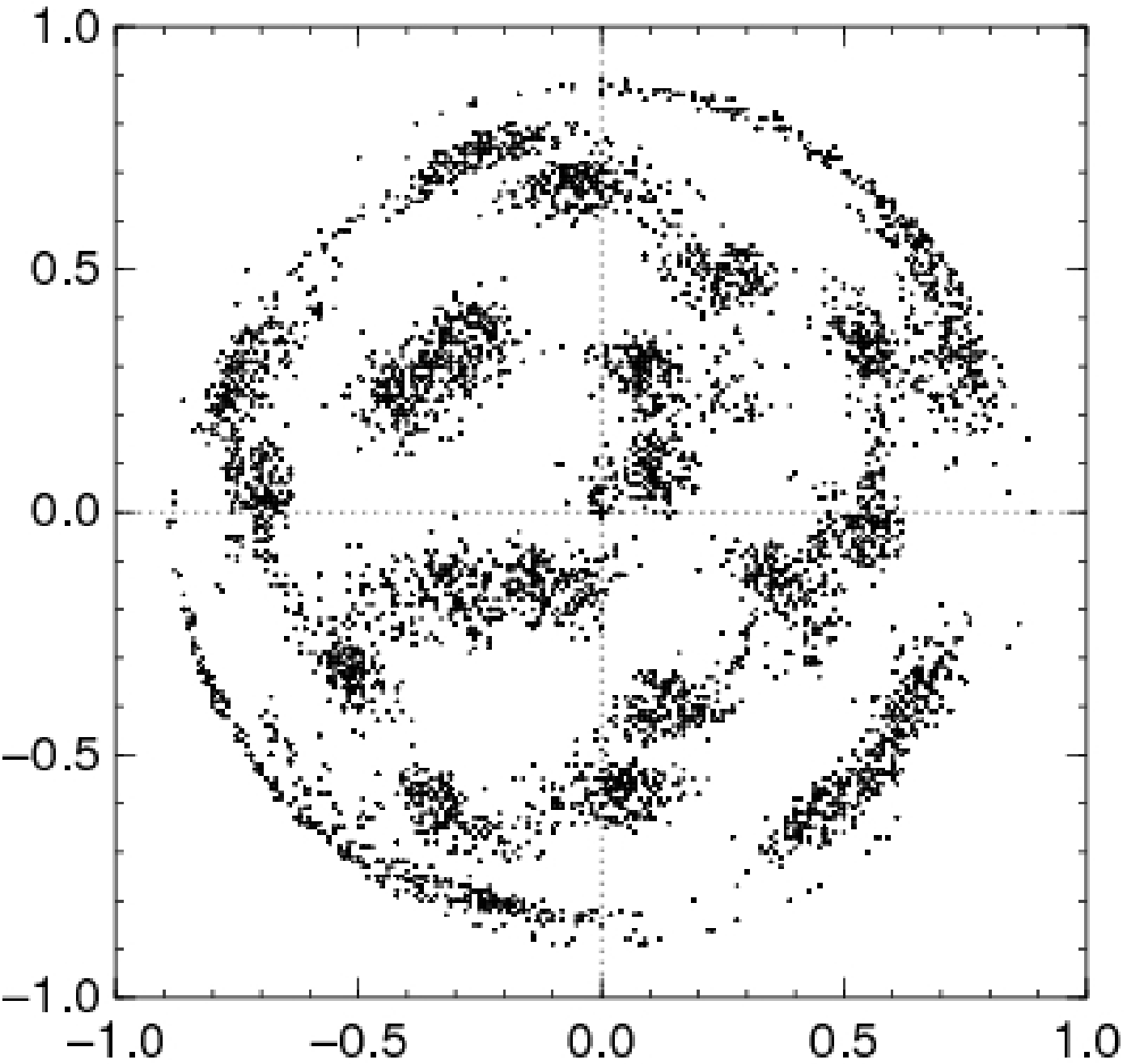}
\caption[]{ \mycaptionfont Full sky map 
  showing the optimal
  orientation of dodecahedral face centres based on 
  40,000 steps in
  4 MCMC chains, using the two simulated three-year INC maps
  with \postrefereechanges{the} 
  highest $\alpha$ estimates (simulations 58 and 80, in the upper and
  lower panels respectively),
  and the kp2 mask, 
  showing face centres
  for which $P > 0.5$.
}
\label{f-INCthree_lbth_N_sim_hiS}
\end{figure} 
} 

\newcommand\falphaphi{
\begin{figure}
\centering 
\includegraphics[width=8cm,bb=53 173 489 570]{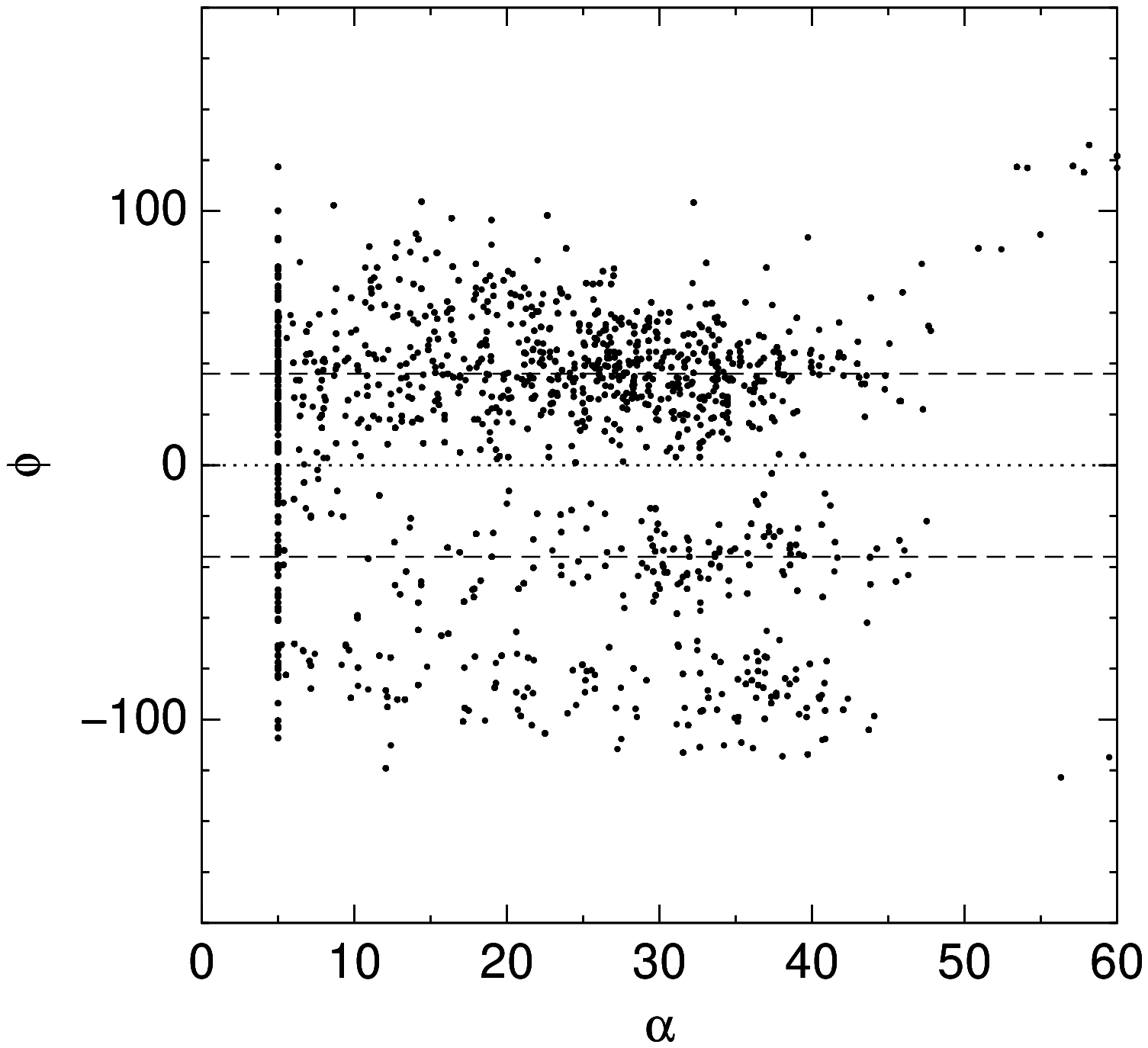} 
\includegraphics[width=8cm,bb=53 173 489 570]{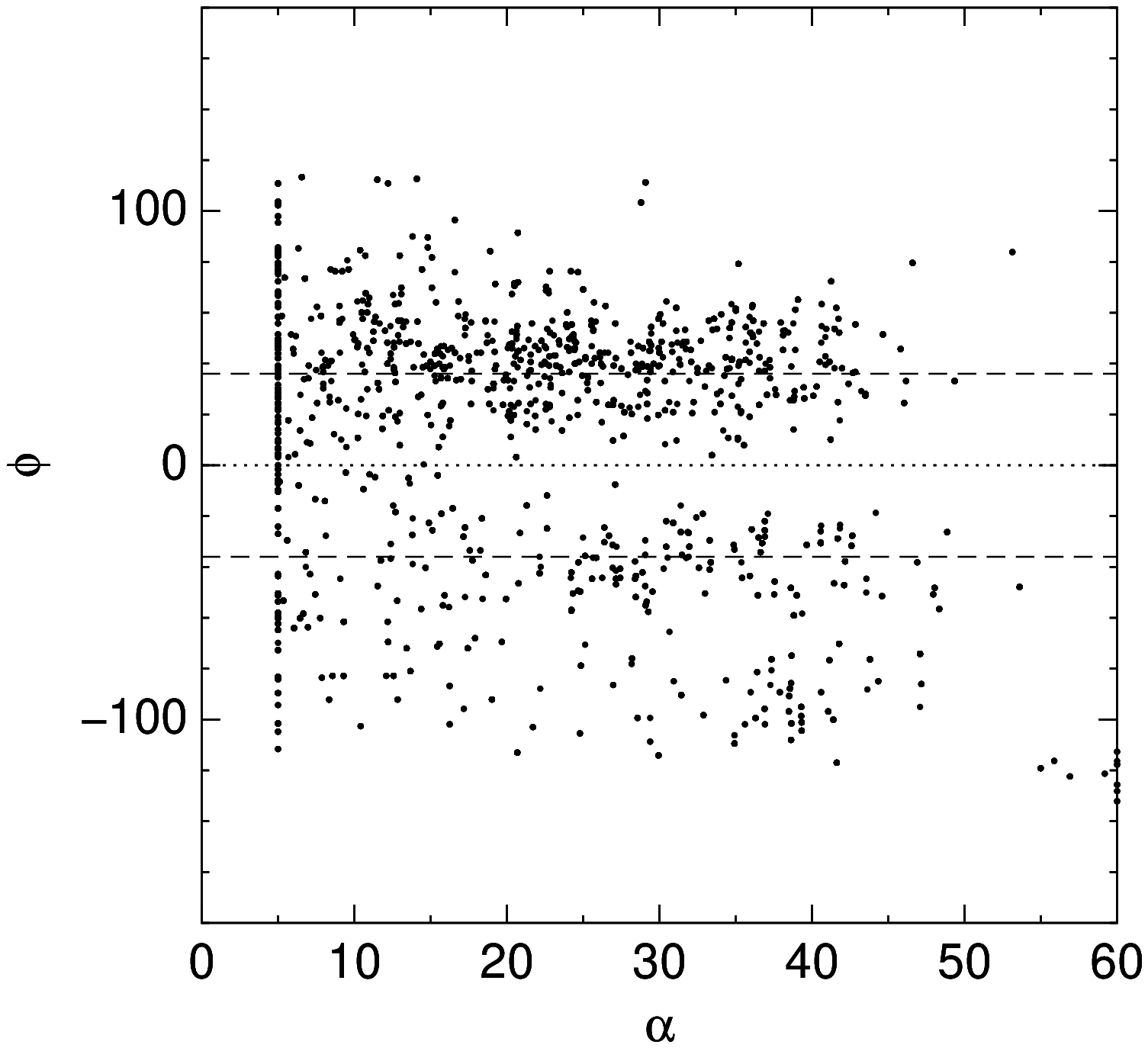} 
\caption[]{ \mycaptionfont
Distribution of $\alpha$ and $\phi$ states where 
$P > 0.5$ in the MCMC
chains of the dodecahedral solution used 
in Table~\protect\ref{t-dodec}, for the ILC (upper panel)
and TOH (lower panel) maps. 
\protect\postrefereechanges{In this and similar figures, lines indicating
$\pm 36\ddeg$ are shown. These are not fit to the data.}
}
\label{f-alpha-phi}
\end{figure} 
} 

\newcommand\fphihist{
\begin{figure}
\centering 
\includegraphics[width=6cm]{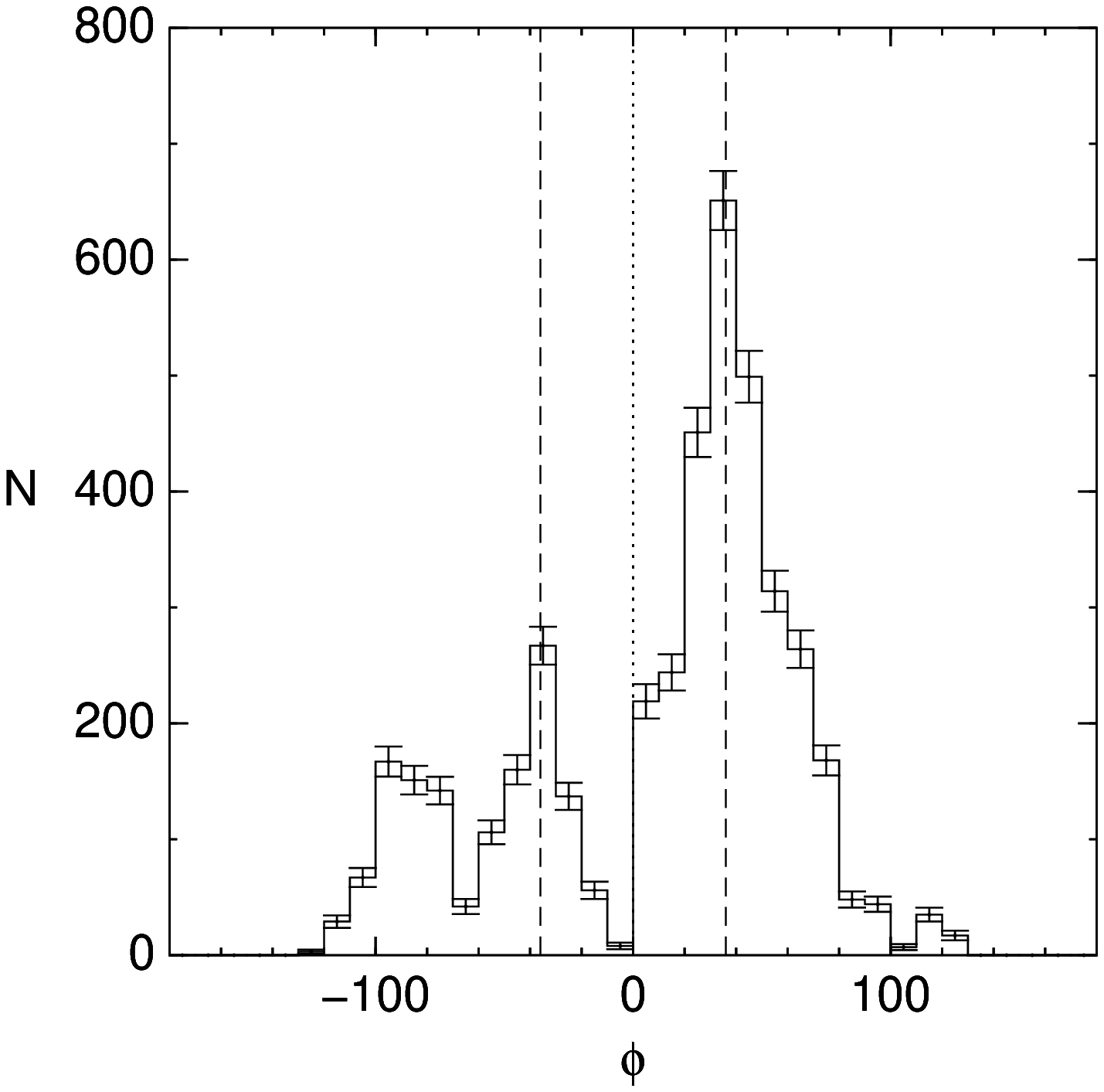} \\
\includegraphics[width=6cm]{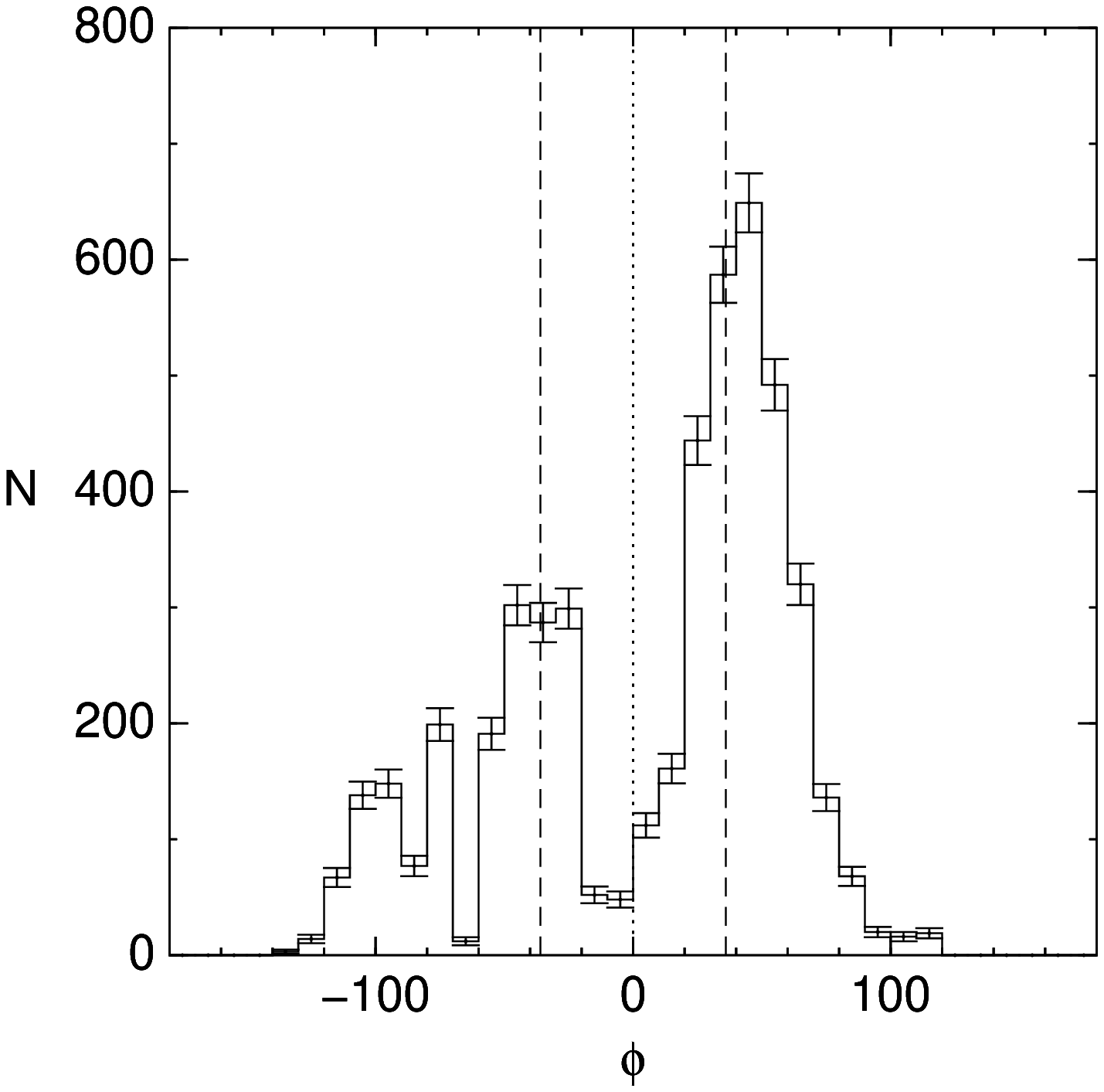}
\caption[]{ \mycaptionfont
\protect\postrefereechanges{Histogram of the twist angle $\phi$ 
shown in Fig.~\protect\ref{f-alpha-phi}, 
with $\sqrt{N}$ error bars, for
the ILC (upper panel)
and TOH (lower panel), excluding states with $\alpha \le 15\ddeg$.}
}
\label{f-phi_hist}
\end{figure} 
}  

\newcommand\fphihistINC{
\begin{figure}
\centering 
\includegraphics[width=6cm]{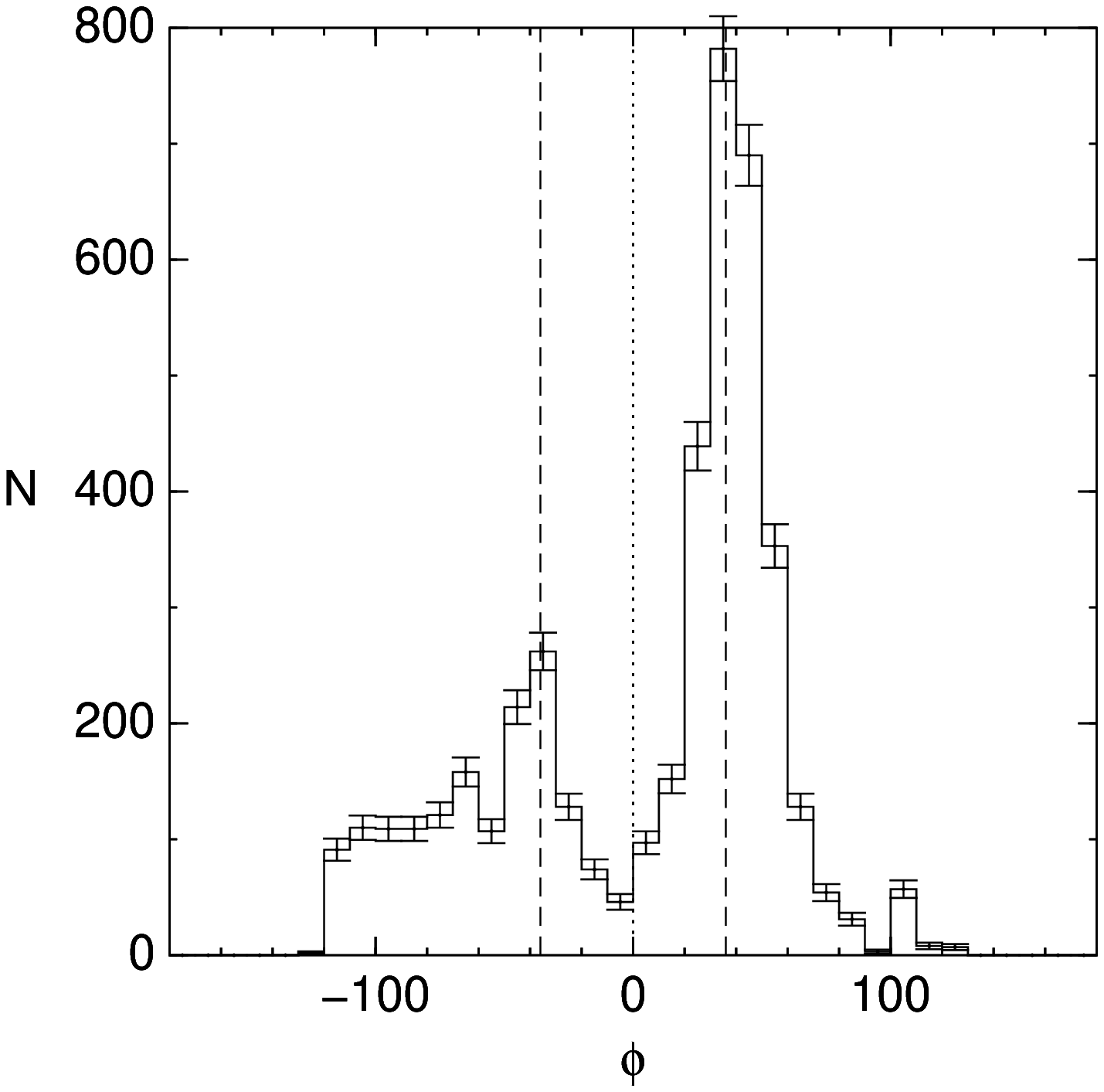} 
\caption[]{ \mycaptionfont
\protect\postrefereechanges{Distribution of the twist angle $\phi$ 
in 16 MCMC chains run on the INC3 observational map, for states where $P > 0.5$
(the four chains shown in Fig.~\protect\ref{f-alpha-phi-inc3} plus
12 additional MCMC chains),
with $\sqrt{N}$ error bars, excluding states with $\alpha \le 15\ddeg$.}
}
\label{f-phi_hist_INC3}
\end{figure} 
}  

\newcommand\falphaphiINCthree{
\begin{figure}
\centering 
\includegraphics[width=8cm,bb=53 173 489 570]{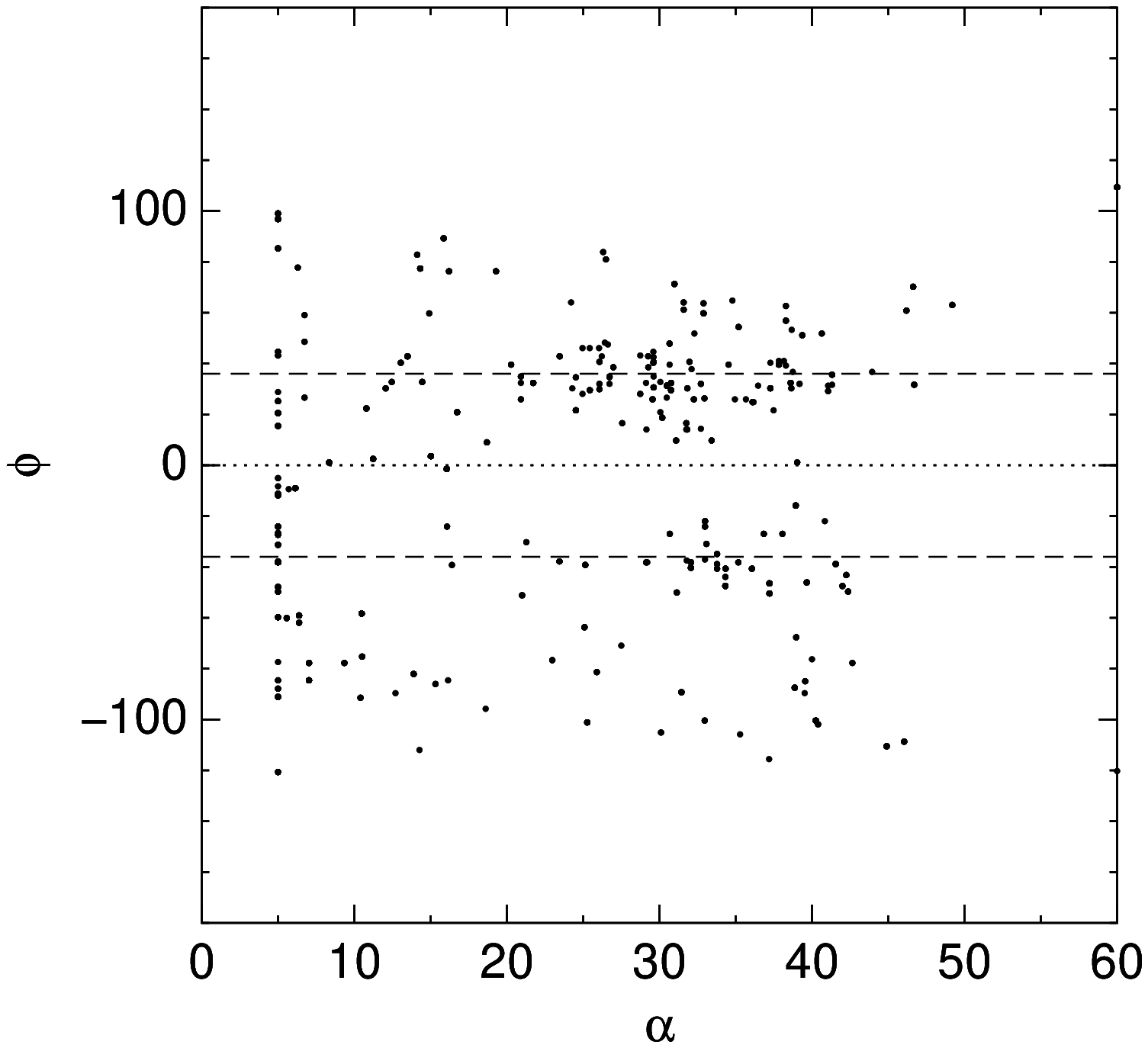} 
\caption[]{ \mycaptionfont
Distribution of $\alpha$ and $\phi$ states where 
$P >  0.5$ in the MCMC
chains for the INC3 observational map.
}
\label{f-alpha-phi-inc3}
\end{figure} 
} 

\newcommand\falphaphiINCthreesim{
\begin{figure}
\centering 
\includegraphics[width=8cm,bb=53 173 489 570]{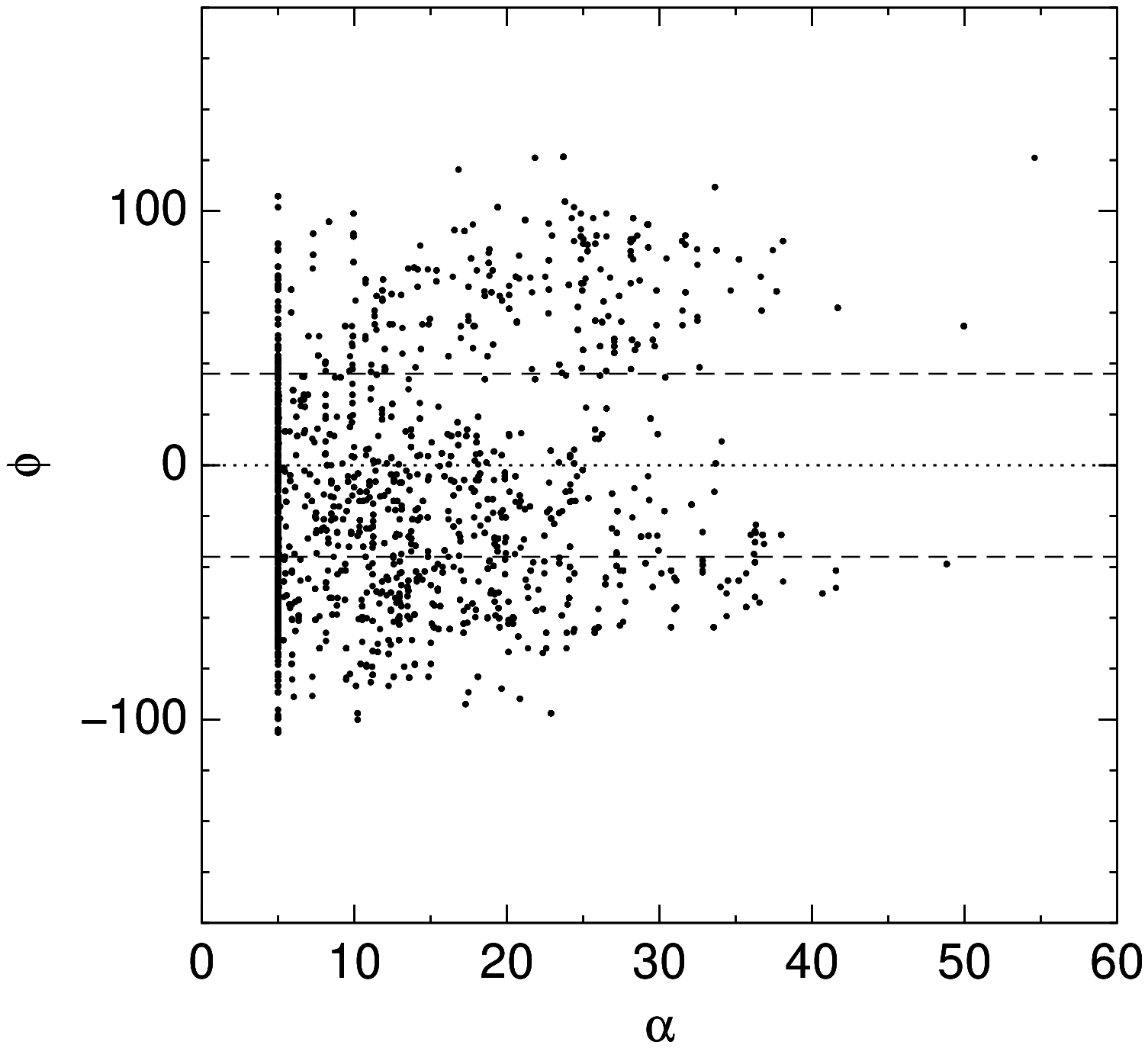} 
\includegraphics[width=8cm,bb=53 173 489 570]{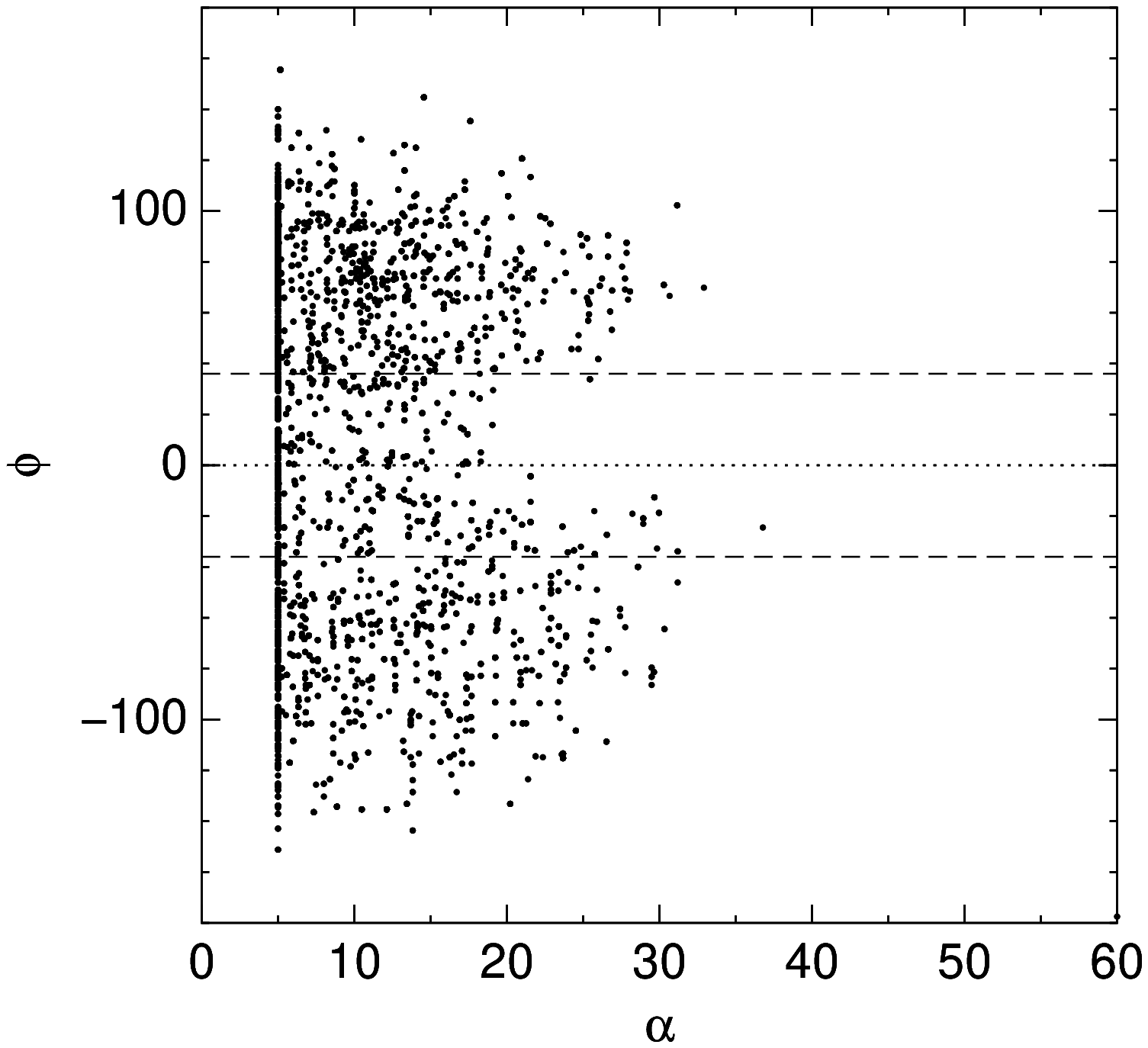} 
\caption[]{ \mycaptionfont
  Distribution of $\alpha$ and $\phi$ states where 
  $P >  0.5$ in the MCMC
  chains for the two simulated three-year INC maps
  with lowest $\Sxi$ estimates (simulations 92 and 90, in the upper and
  lower panels respectively).
}
\label{f-alpha-phi-inc3_sim}
\end{figure} 
} 

\newcommand\falphaphiINCthreesimhiS{
\begin{figure}
\centering 
\includegraphics[width=8cm,bb=53 173 489 570]{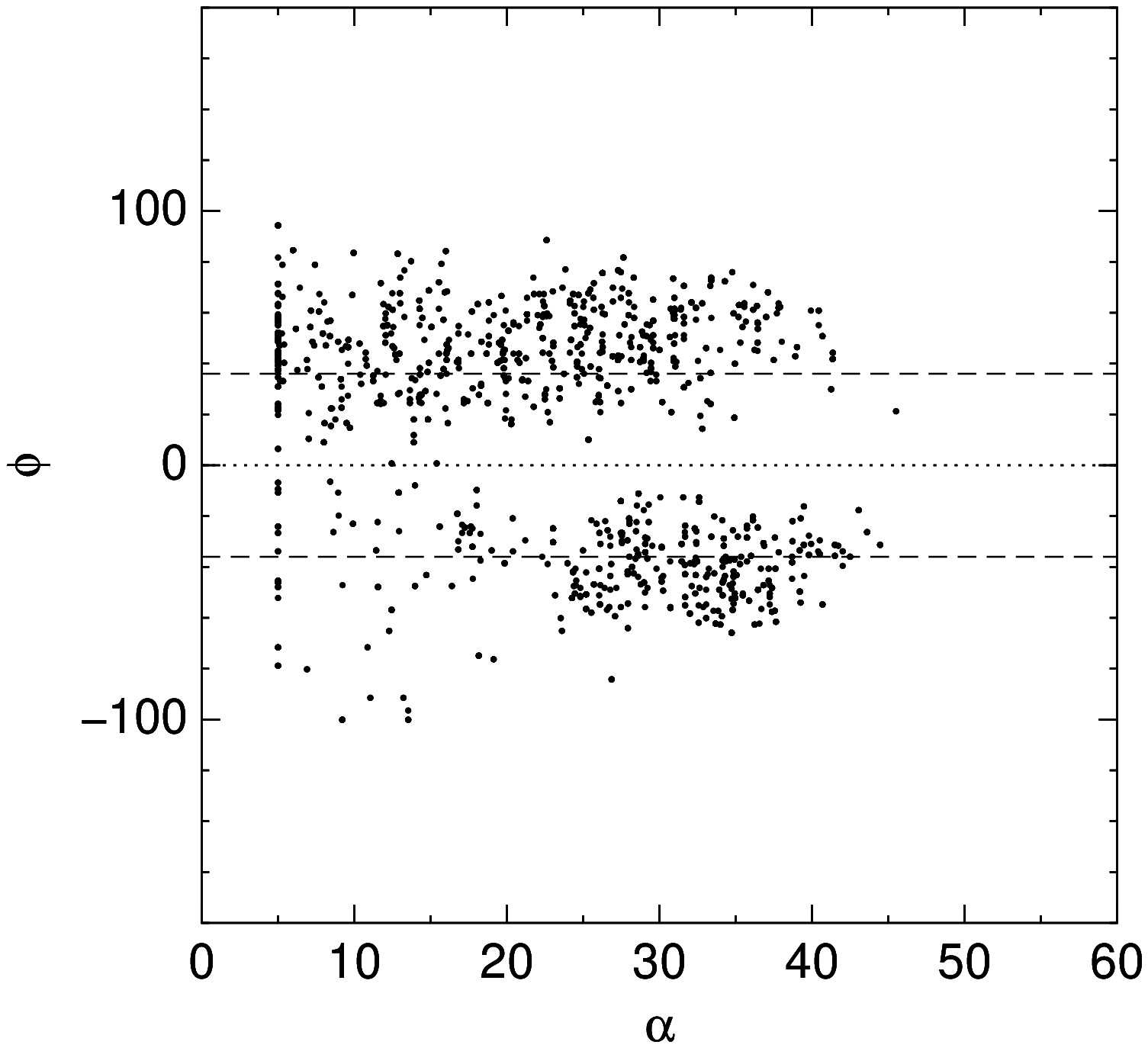} 
\includegraphics[width=8cm,bb=53 173 489 570]{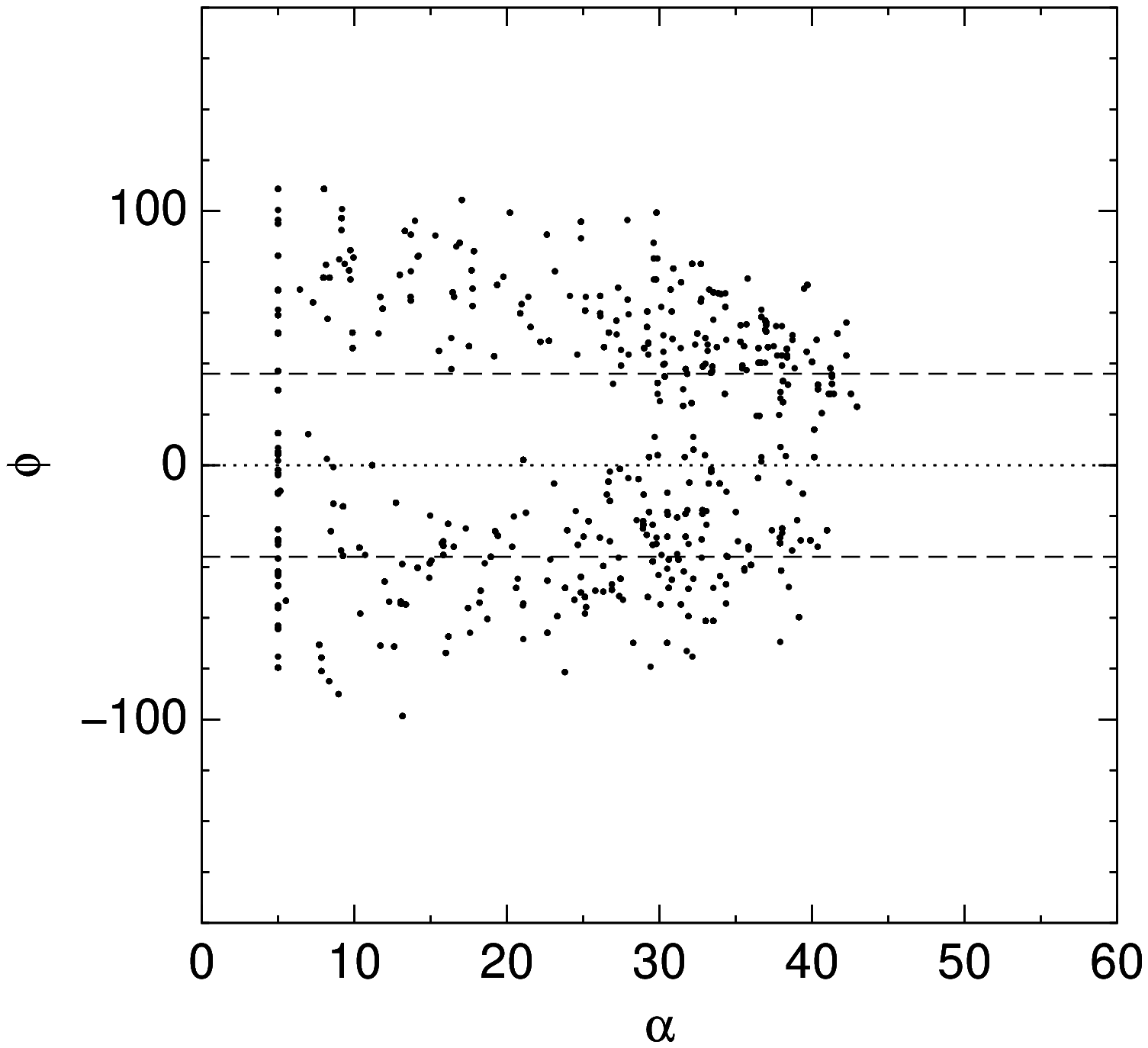} 
\caption[]{ \mycaptionfont
  Distribution of $\alpha$ and $\phi$ states where 
  $P >  0.5$ in the MCMC
  chains for the two simulated three-year INC maps
  with highest $\alpha$ estimates (simulations 58 and 80, in the upper and
  lower panels respectively).
}
\label{f-alpha-phi-inc3_sim_hiS}
\end{figure} 
} 

\newcommand\fsymsig{
\begin{figure}
\centering 
\includegraphics[width=6cm]{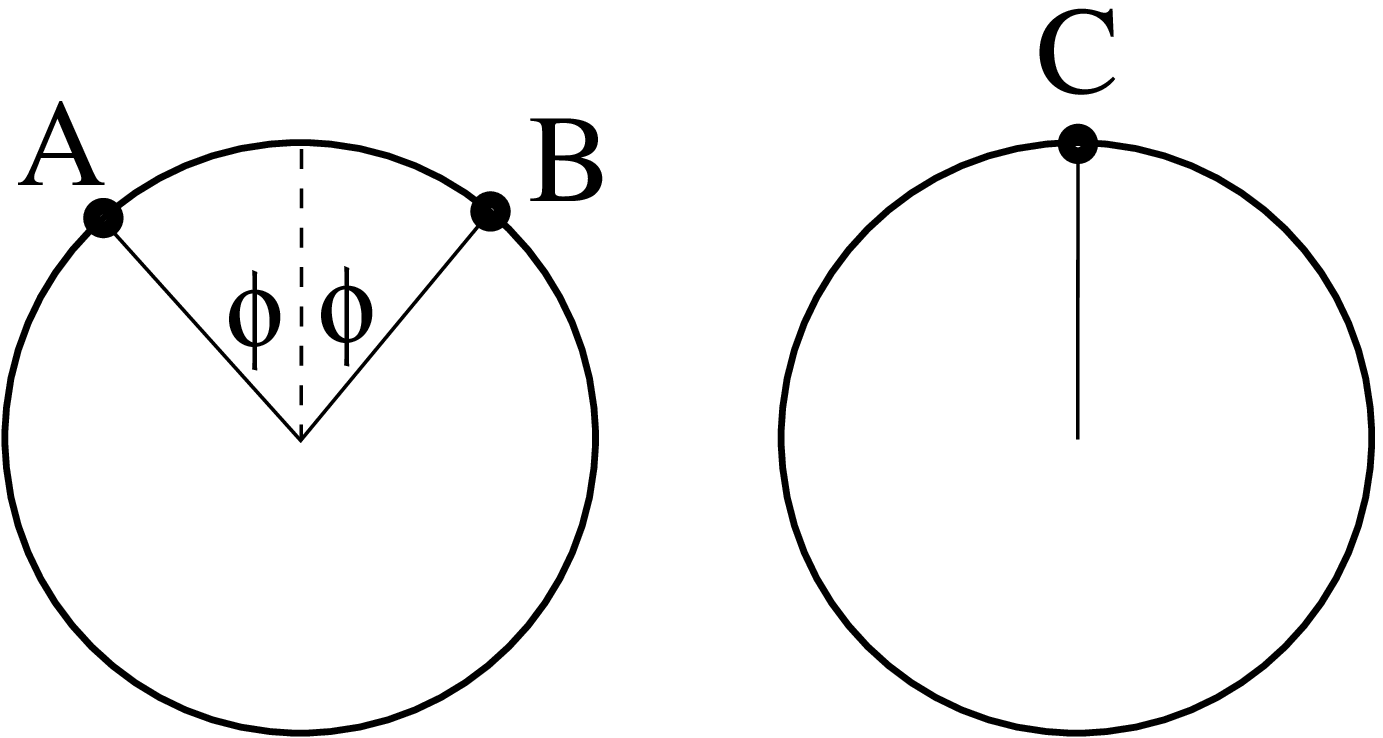}
\caption[]{ \mycaptionfont Schematic diagram showing two circles on
  the SLS on which cross-correlation is to be optimised, for 
  twist $\phi =0$.
  The fluctuations are zero everywhere on the two
  circles except at the points A and B on the first circle and C on
  the second circle.  At all three points the fluctuation amplitude is
  $+1$.  }
\label{f-symsig}
\end{figure} 
} 

\newcommand\fantipode{
\begin{figure}
\centering 
\includegraphics[width=6cm]{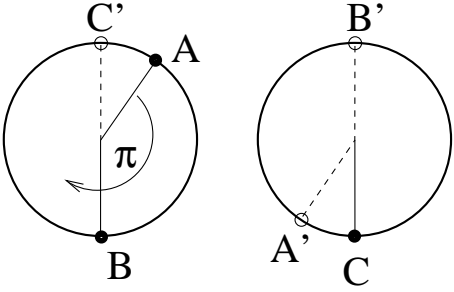} 
\caption[]{ \mycaptionfont Schematic diagram 
  of two circles on the SLS for 
  twist $\phi =0$, illustrating that the 
  observed anti-correlation at antipodal points on the SLS corresponds
  to an anti-correlation at twist $\phi = \pi$ on the
  circles. Solid circles at points A, B, and C
  indicate positive fluctuations, and 
  hollow circles at the antipodal points A', B', and C' indicate negative
  fluctuations.
 }
\label{f-antipode}
\end{figure} 
} 

\newcommand\falphaphiantip{
\begin{figure}
\centering 
\includegraphics[width=8cm]{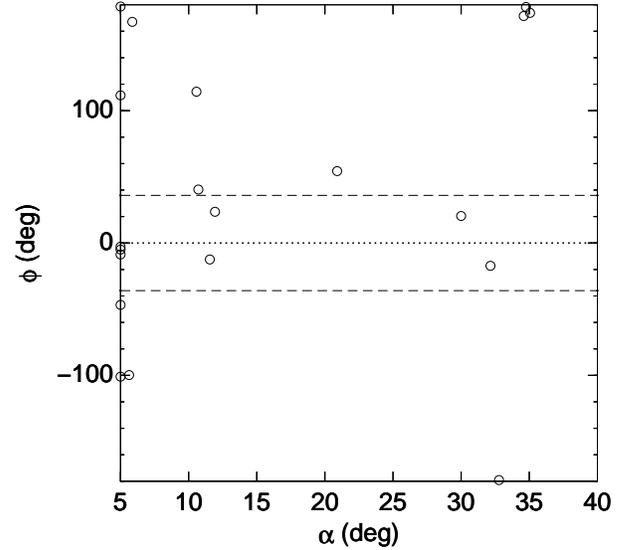}
\caption[]{ \mycaptionfont
Median circle sizes $\alpha$ and twist angles $\phi$ (mean) 
for each of the 20 simulations, 
\postrefereechanges{analysed using the steps with $P > 0.5$,}
modified as described
in \SSS\ref{s-why-phi-distbn} by inverting the anti-correlation in 
50 randomly selected pairs of 10$\ddeg$ radius antipodal patches on
the sky.
}
\label{f-alpha_phi_antip}
\end{figure} 
} 

\newcommand\fphihistantip{
\begin{figure}
\centering 
\includegraphics[width=6cm]{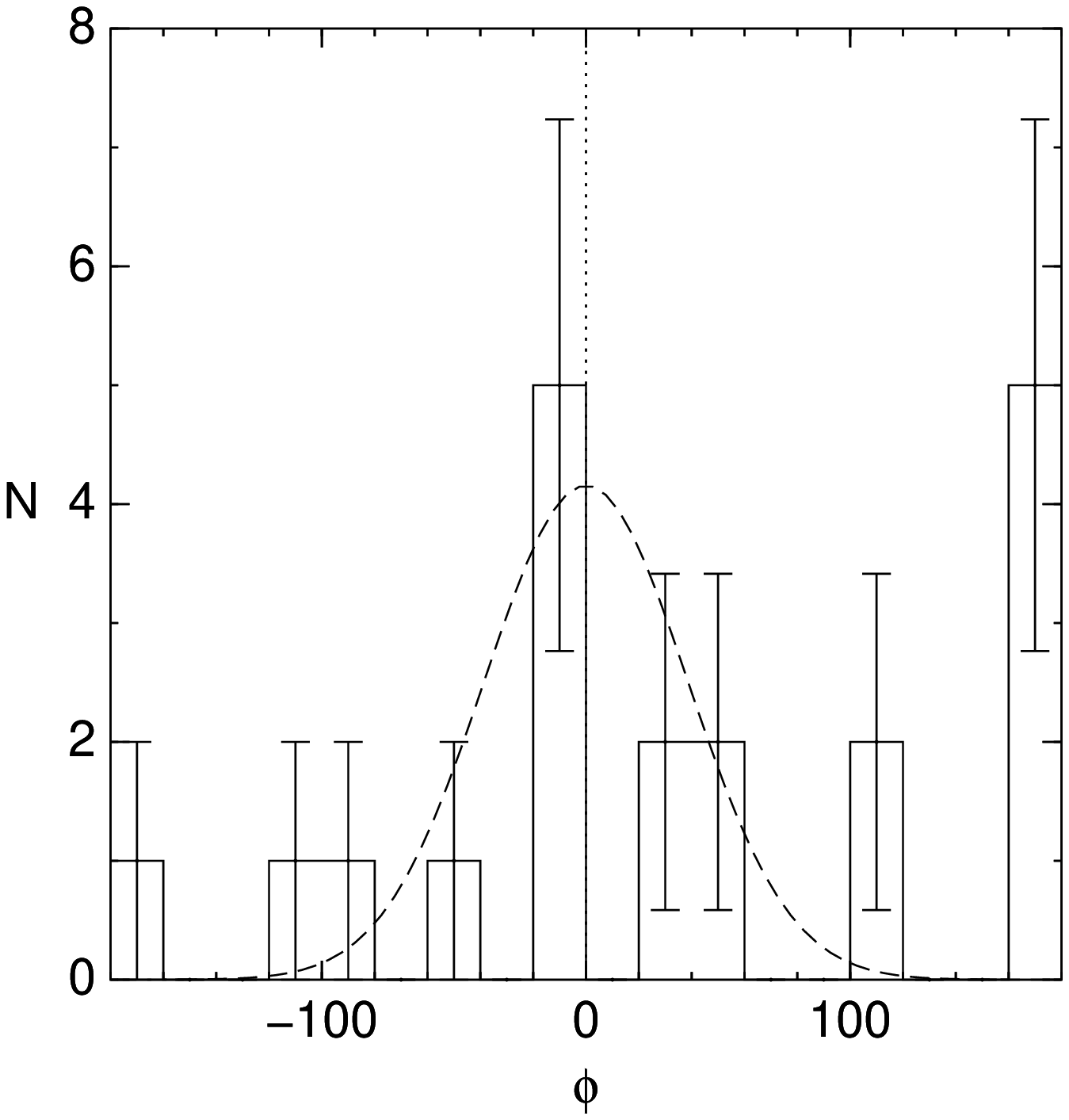}
\caption[]{ \mycaptionfont
\postrefereechanges{Histogram 
of $\phi$ values shown in Fig.~\protect\ref{f-alpha_phi_antip}.
The Gaussian distribution shown in Fig.~\protect\ref{f-phi_hist_sims}
is reproduced here.}
}
\label{f-phi_antip_hist}
\end{figure} 
} 


\section{Introduction}  \label{s-intro}

During the past half-decade, attention has focussed on the Poincar\'e 
Dodecahedral Space (PDS) as a potentially better model of comoving space than
the infinite flat model
\nocite{LumNat03,RLCMB04,Aurich2005a,Aurich2005b,Gundermann2005,KeyCSS06,NJ07,Caillerie07,LR08,RBSG08}({Luminet} {et~al.} 2003; {Roukema} {et~al.} 2004; {Aurich} {et~al.} 2005a, 2005b; {Gundermann} 2005; {Key} {et~al.} 2007; {Niarchou} \& {Jaffe} 2007; {Caillerie} {et~al.} 2007; {Lew} \& {Roukema} 2008; {Roukema} {et~al.} 2008). If the hypothesis that the comoving 
spatial section of the Universe 
is a PDS is correct, then it should be possible to estimate the 
astronomical coordinates of the fundamental domain, as has been tried
by \nocite{RBSG08}{Roukema} {et~al.} (2008) (hereafter, RBSG08). 
Moreover, successive improvements in data and 
analysis methods should yield successively more precise and more accurate
estimates of these astronomical coordinates. 

Here, we aim to improve on the method of optimising the cross-correlation 
$\ximc$ 
\postrefereechanges{[Eq.~(1), RBSG08]} 
of cosmic microwave background (CMB) temperature fluctuations between
copies of the surface of last scattering (SLS) presented in \nocite{RBSG08}{Roukema} {et~al.} (2008) 
and also applied by \nocite{Aurich08a}{Aurich} (2008).
We investigate whether or not this can lead to improved parameter estimates.
We also test a small number of simulations
to see whether the infinite
flat model can reproduce the observational signal.

In RBSG08, the method was introduced by pointing out that algebraically,
it is an extension of using the identified circles principle first
published by \nocite{Corn96,Corn98b}{Cornish} {et~al.} (1996, 1998). 
Here, we make use of the fact that 
the cross-correlation method is not only an algebraic extension of the 
identified circles principle, but its relation to the identified 
circles principle can also be interpreted in terms of identified {\em annuli}.
This leads to algebraic/trigonometric relations
that enable a faster calculation of cross-correlations, without
sacrificing the uniformity of the underlying random selection of points
on the sky. Moreover,
use of these relations also increases the numbers of pairs per separation
bin, leading to less relative Poisson noise in individual cross-correlation 
estimates.
The annulus interpretation and the associated relations
are presented in \SSS\ref{s-method-anglimits}.

In \SSS\ref{s-method-fiveyr}, use of this faster method
to recalculate Monte Carlo Markov Chains (MCMC) 
for the five-year release of the WMAP data is described. 
Since the new method increases the numbers of pairs per separation bin,
the smallest separation bin available when estimating correlations 
should not be as noisy as without this new method, so we include this
smallest separation bin.

Use of this faster method also makes it practical to perform 
a consistent analysis of both the WMAP map and a small
number of Gaussian random fluctuation (GRF) WMAP simulated maps, in order to
estimate the expectation of a PDS-like signal under the assumption of
a simply connected model. 
In order to model this, we need to consider the following.

One of the key motivations from WMAP data for studying multiply connected models has
been the lack of structure on angular scales larger than about 60{\ddeg},
as has been noted by several authors 
\nocite{WMAPSpergel,LumNat03,Aurich2005b}(e.g.  {Spergel} {et~al.} 2003; {Luminet} {et~al.} 2003; {Aurich} {et~al.} 2005b). 
As was 
noted in RBSG08, in particular in Eq.~(10), if the PDS model
is incorrect, then cross-correlations of temperature fluctuations 
at pairs of points that are implied (by the PDS hypothesis) to be spatially 
close should, on average, be much smaller than the auto-correlations at
the same small spatial separation scale. This is because if the PDS model
is wrong, then these pairs are in reality widely separated rather than closely
separated, so their correlations should, on average, be small.

However, this is only a statistical statement: even for the case that
the PDS model is physically wrong, it is possible that some
orientations of a PDS model, by chance, happen to give cross-correlations
that violate Eq.~(10) of RBSG08. 
How probable 
is it for chance large-scale correlations in particular
directions to mimic a PDS-like signal? This clearly depends on the
amplitude of the large length-scale auto-correlation. If this
amplitude is low or high, then the chance of finding PDS-like signals using the
optimal cross-correlation method should be low or high
respectively. Since the range of matched circle sizes $\alpha$ studied in 
RBSG08 included circles with
$ \alpha \le 60\ddeg$, most of the cross-correlations are from pairs
of points separated observationally by angles $\gtapprox 60\ddeg \sim 
1\; \mbox{rad}$.

The latter is the minimum scale above which \nocite{WMAPSpergel}{Spergel} {et~al.} (2003)
quantified the lack of temperature-temperature auto-correlations at
large scales in the WMAP data, with their parameter $S$, 
\postrefereechanges{defined
\begin{equation}
S = \int^{1/2}_{-1} [C(\theta)]^2 \mathrm{d} \cos \theta,
\label{e-Sxi-spergel}
\end{equation}
where $C(\theta)$ is the two-point auto-correlation function at 
an angular separation of $\theta$ on the SLS}
[Sect.~7, Eq.~(9), \nocite{WMAPSpergel}{Spergel} {et~al.} 2003]. 
\postrefereechanges{Hereafter}, we call this $\Sxi$, in order
to distinguish it from the $S$ parameter used for matched circle
analyses. The two parameters have very different meanings.

The chance of observing $\Sxi$ to be as small as that observed was estimated
by \nocite{WMAPSpergel}{Spergel} {et~al.} (2003) as 0.15\% for an infinite, flat, cosmic concordance model 
with a fixed spectral index of density perturbations.
\nocite{EfstNoProb03b}{Efstathiou} (2004), using what should be a more accurate method if we  
assume a simply connected model, 
estimated the chances to be much higher, 
from 3\% to 12.5\% depending on which sky map is analysed and which (if any)
galactic cut mask is used (Table~5, \nocite{EfstNoProb03b}{Efstathiou} 2004).

We can quantify the relation between large length-scale auto-correlations
and the chance of a PDS signal in the case that the PDS is wrong by writing
that {\em the lower the observed value of $\Sxi$, the less likely it
is that a non-PDS model will give a PDS-like signal.}
Hence, since we expect that
the two properties, a low $\Sxi$ value and a low chance of a PDS-like
cross-correlation signal occurring in a flat, infinite, cosmic
concordance model, are related, it would not be useful to estimate
their probabilities independently.
Instead, 
what is of interest to investigate is
the chance that a PDS-like cross-correlation signal occurs, {\em given that
$\Sxi$ is as low as that which is observed.} Conservatively, we can 
calculate an upper limit to this conditional probability if we use
simulated sky maps with $\Sxi$ values a little higher than the observed
estimate of $\Sxi$.  Moreover, here we use the kp2 galactic cut mask,
for which \nocite{EfstNoProb03b}{Efstathiou} (2004) finds the
highest probability for the observed $\Sxi$ value to occur in an
infinite, flat, cosmic concordance model, so we should obtain a
conservative upper limit to the probability of both (a) an $\Sxi$ value
as low as that observed and (b) a PDS-like cross-correlation signal
similar to that observed occurring.
This is described in more detail in \SSS\ref{s-method-GRF}.



Results are presented in \SSS\ref{s-results}, discussed in \SSS\ref{s-disc}
and conclusions are given in \SSS\ref{s-conclu}.
For general background on spherical, multiply connected spaces, 
see \nocite{Weeks2001}{Weeks} (2001),
\nocite{GausSph01}{Gausmann} {et~al.} (2001), 
\nocite{LehSph02}{Lehoucq} {et~al.} (2002) and
\nocite{RiazSph03}{Riazuelo} {et~al.} (2004). See the references cited above for background 
on cosmic topology. Comoving coordinates are used when discussing distances 
(i.e. ``proper distances'' at the present epoch, \nocite{Wein72}{Weinberg} 1972, 
equivalent to ``conformal time''
if $c=1$) and the Hubble constant is written $H_0 \equiv 100 h$\kms/Mpc.


\section{Observations and simulations} 

\subsection{Observations} \label{s-wmap}

The analysis described in \SSS\ref{s-method-fiveyr} 
uses the Internal Linear Combination 
(ILC)\footnote{\url{http://lambda.gsfc.nasa.gov/data/map/dr3/dfp/} 
\url{wmap_ilc_5yr_v3.fits}}
all-sky map of the 
five-year WMAP data \nocite{WMAP5Hinshaw}({Hinshaw} {et~al.} 2008)
and the foreground cleaned, Wiener filtered 
version of the same five-year data published by the
\nocite{WMAPTegmarkFor}{Tegmark} {et~al.} (2003) group 
(TOH)\footnote{\url{http://space.mit.edu/home/tegmark/wmap/} 
\url{wiener5yr_map.fits}}
The ``kp2'' mask
to eliminate the Galactic Plane and other likely 
contaminating regions, \postrefereechanges{covering} about 15\% of the sky,
is used \postrefereechanges{throughout this paper unless otherwise 
noted}.\footnote{Data file:
\url{http://lambda.gsfc.nasa.gov/data/map/dr2/}
\url{ancillary/wmap_kp2_r9_mask_3yr_v2.fits};  \\
map projection: 
\url{http://lambda.gsfc.nasa.gov/product/map/}
\url{current/map_images/f02_int_mask_b.png}.}
We do not smooth these maps.


\subsection{Simulations and associated observational map}
\label{s-inc-sims}

The analysis described in \SSS\ref{s-method-GRF} uses a version of
the WMAP map and sky simulations prepared consistently from
three-year observational and simulated maps in the Q, V and W frequency 
bands, weighted by inverse noise as given in \nocite{Hinshaw03}{Hinshaw} {et~al.} (2003),  
and smoothed by a Gaussian of FWHM {1\ddeg}, 
as described in Sect.~2 of \nocite{LR08}{Lew} \& {Roukema} (2008).\footnote{The maps can be
downloaded from 
\protect\url{http://cosmo.torun.pl/GPLdownload/MCMC/}
\protect\url{sims-from-LR08-project/}.}
This version of the observational map is referred to hereafter as 
``INC3''.\footnote{Inverse Noise Coadded 3-year}

As mentioned above, the expected value of $\Sxi$ from the infinite,
flat, cosmic concordance model is higher than that observed.  However,
there is a (small) chance in the model that such a low $\Sxi$ value
occurs in any single realisation of the model, 
such as the one in which we live.
Since we wish to test the chance that a PDS-like
cross-correlation signal occurs {\em given that $\Sxi$ is as low as that
which is observed,} we generate simulations using the observational
estimates of the  spherical harmonic
amplitudes $C_l$ of the 
temperature fluctuations as estimated in 
\nocite{Hinshaw06}{Hinshaw} {et~al.} (2007), rather than using the mean values implied by an infinite
flat model. The phases of the spherical
harmonics are randomised. Gaussian noise is simulated according to the
properties and scanning strategy of each differencing assembly and
added to each simulated map.

Since a lot of the large-scale power in what are considered as the 
best estimates of the cosmological signal in the WMAP data lie close to
the Galactic Plane \nocite{WMAPTegmarkFor}(e.g.  {Tegmark} {et~al.} 2003), simulations based
on spherical harmonics with the same $C_l$'s as this signal but different
phases will most of the time have high large-scale power which does not
happen to lie inside the kp2 cut. Hence, simulations made with the 
\nocite{Hinshaw06}{Hinshaw} {et~al.} (2007) $C_l$ estimates and masked with the kp2 mask will,
in general, have $\Sxi$ estimates {\em larger} than that which is actually
observed for the part of the signal outside of the kp2 cut rather than
approximately equal to it. 

This is not due to an error in the \nocite{Hinshaw06}{Hinshaw} {et~al.} (2007) estimates 
(assuming a simply connected model\footnote{For a multiply connected model,
the distributions of the various $a_{lm}$'s are to some degree dependent
on one other, so a method which assumes they are independent can at 
best give an approximate result.}).
Even though these estimates for
$l\le30$ are made using nearly the same kp2 cut that we use here 
(\nocite{Hinshaw06}{Hinshaw} {et~al.} 2007 use a kp2 cut degraded in resolution), the
spherical harmonics are nevertheless functions of the full
sky. \nocite{Hinshaw06}{Hinshaw} {et~al.} (2007) use a maximum likelihood method which uses the
information from the cut sky to infer the best estimate of the $C_l$
values for these functions covering the full sky. Unsurprisingly, it
implicitly extrapolates from the cut sky to the full sky, and recovers
some of the information masked in the Galactic Plane.

Since we wish to test the chance that a PDS-like
cross-correlation signal occurs given that $\Sxi$ is as low as that
which is observed, we have two obvious approaches to choose from in order
to have simulations which are statistically comparable with the observations.
One approach would be to analyse the full sky for both observations and simulations. 
In this case, the $\Sxi$ estimates should be approximately equal. However,
a large part of the signal would be that from inside the Galactic Plane. 
The risk of implicitly testing the properties of foreground contaminants 
rather than of the cosmological signal would be high. 

The other approach is to analyse the cut sky (with the kp2 mask) for
both observations and simulations. This decreases the risk that our
analysis will be affected by contamination from the Galactic
Plane. However, for the cut sky, many of the simulations will have $\Sxi$ 
higher than $\Sxi$  of the observations.
In this case, it will be necessary 
to select those simulations with the lowest values of $\Sxi$
not greater than $\Sxi$ of the observational map. 

This will not give an exact
statistical match between simulations and observations. 
However, since greater $\Sxi$ should increase the chance of cross-correlations
occurring, this should give an upper limit for estimating the probability of
cross-correlations occurring in a simply connected model, i.e. a conservative
estimate. This is the approach we adopt here.

We estimate $\Sxi$ for each of the observational and simulational using
\begin{equation}
\Sxi \equiv \int_{\rSLS}^{2\rSLS} [\xisc(r)]^2 \; \mbox{d} \frac{r}{\rSLS},
\label{e-defn-Sxi}
\end{equation}
where $\rSLS$ is the radius of the SLS, $\xisc(r)$ is the 
(comoving) spatial auto-correlation function
as given in Eq.~(4) of RBSG08, 
\postrefereechanges{and $r$ is the 
comoving spatial separation of a pair of points on the SLS.}
 This integral differs 
from
that in Eq.~(9) of \nocite{WMAPSpergel}{Spergel} {et~al.} (2003), since here we focus on spatial 
separation, 
while \nocite{WMAPSpergel}{Spergel} {et~al.} (2003) use an orthogonal
projection of the angle between a pair of sky positions. 
The minimum angular separation used by the latter is 
$60\ddeg \approx 1\, \mathrm{rad}$, so here we use $\rSLS$ for the minimum separation.
Provided that we estimate $\Sxi$ in the same way for both observational data
and simulations, the difference between spatial and projected angular 
definitions of $\Sxi$ should not affect the questions of interest here.


\falphap

\falpham


\section{Method} \label{s-method}


The method of using Markov Chain Monte Carlo simulations to optimise
the 
\postrefereechanges{cross-correlation $\ximc$} 
of temperature fluctuations between copies of
the surface of last scattering (SLS) in the covering space $S^3$,
modelled as $S^3 \subset {\mathbb R}^4$,
is described in Sect.~3 of RBSG08.

Since it is important that the choice of pairs of points is as uniform
as the nature of the observations allows, the selection of points on 
a copy of the SLS in RBSG08 
was chosen uniformly on the SLS, i.e. on the 2-sphere.
The binning into bins of pair separations was calculated only after 
the pair separations have been calculated by applying a holonomy 
transformation to one of the members of the pair.

From a computational point of view, applying the holonomy is the most
intensive step in the calculation, requiring the multiplication of a
$4\times4$ matrix by a 4-vector, as a rotation in ${\mathbb R}^4$.
This matrix calculation is carried out 12 times for a given pair of
points, since all of the 12 holonomy transformations mapping a point
to one of its neigbouring copies of the SLS must be examined.

\postrefereechanges{The set of 20 holonomy transformations that maps
  the fundamental domain to the next layer in the direction of the
  hyperspherical equator could, in principle, be used too. However,
  for these to give matched circles as small as $5\ddeg$, this would
  require the total matter-energy density to be $\Omtot \gtapprox
  1.03$, which is uncomfortably high given present observational
  estimates of $\Omtot$. To see this, replace $\pi/10$ in Eq.~(15) of
  RBSG08 by $\pi/6$, which is half the geodesic length of any member
  of this set of holonomy transformations (Clifford translations), and
  use $\rC = (c/H_0) (\Omtot-1)^{-1/2}$, where $\rC$ is the radius
  of curvature, i.e. the radius of $S^3$ modelled as a subset of
  ${\mathbb R}^4$.  Moreover, it is not clear how a ``generalised''
  twist parameter could be used if both sets (or yet further sets) of
  holonomy transformations were to be used in a single estimate of the
  cross-correlation function. The twist for this set of holonomy
  transformations is $\pm \pi/3$, not $\pm \pi/5.$ For these reasons,
  we consider just the 12 immediately neighbouring copies of the
  fundamental domain, as in RBSG08.}

\postrefereechanges{In this paper}, we argue that after uniformly
selecting points on the 2-sphere, a filtering of these points is
possible in a way that excludes only those points that are certainly
not members of any close SLS-SLS pairs, for a given maximum pair
separation $r_2$.  This should enable shortening the time to find a
given number of close pairs for a constant number of points $N_p$, as
well as finding a larger number of close pairs for the same value of
$N_p$, since points that are certainly not members of any close pair
are rejected before the iteration over pairs of points is started.
The latter improvement to the algorithm should enable practical use of
the method for shorter pair separations.

\subsection{Preselection of potential members of close pairs: 
$\alpha_1, \alpha_2$}
\label{s-method-anglimits}

Figures~\ref{f-alpha_p} and \ref{f-alpha_m} show how a close pair used in
a cross-correlation estimate relates to the two copies of the SLS and
the angle $\alpha_{\pm}$ that separates it, on a copy of the SLS, from
a dodecahedral face centre. 
Using the 
spherical cosine formula, the lower triangle in the two figures 
has the relation
\begin{equation}
\cos \frac{x}{\rC} = \cos\frac{\rSLS}{\rC} \cos\frac{\pi}{5} + 
         \sin\frac{\rSLS}{\rC} \sin\frac{\pi}{5} \cos \alpha_{\pm}
\end{equation}
where $\alpha_+$ and $\alpha_-$ are for 
Figs~\ref{f-alpha_p} and \ref{f-alpha_m} respectively. 

Clearly, $\alpha_+$ is maximised when $x$ is maximised.
For fixed $r_2$ and $\rSLS$, $x$ 
is maximised when $x=\rSLS + r_2$, i.e.  when the
upper triangle degenerates into a single line segment.
Hence, the angle $\alpha_+$ is maximised when $x=\rSLS + r_2$.

Similarly, $\alpha_-$ is minimised when $x$ is minimised, 
provided that $r_2 \le \rSLS$, which is the situation most interesting
for the SLS-SLS cross-correlation method, since close pairs are the most
useful.
For fixed $r_2$ and $\rSLS$, $x$ 
is minimised when $x=\rSLS - r_2$, i.e.  when the
upper triangle degenerates into a single line segment.
Hence, the angle $\alpha_-$ is minimised when $x=\rSLS - r_2$.

Hence, by symmetry, for a given separation $r_2 \le \rSLS$, the minimum
and maximum boundaries for defining an annulus 
\begin{equation}
\alpha_- \le \alpha \le \alpha_+
\end{equation}
on the SLS around a dodecahedron face centre, in order to include all
points that can potentially be members of a close pair separated by 
a spatial geodesic distance of at most $r_2$ for the corresponding pair
of matched faces of the fundamental domain,
are
\begin{equation}
\alpha_{\pm} = \cos^{-1} \left[ 
 \frac{ \cos\frac{\rSLStiny \pm r_2}{\rC} - \cos\frac{\rSLStiny}{\rC} \cos\frac{\pi}{5} }
 {\sin\frac{\rSLStiny}{\rC} \sin\frac{\pi}{5} } \right] .
\label{e-alpha-pm}
\end{equation}
When $r_2$ is larger than the separation of dodecahedral face centres
(Eq.~(32) in RBSG08), i.e. when
\begin{equation}
r_2 > 2 \rC \left( \frac{\rSLS}{\rC} - \frac{\pi}{10} \right) ,
\label{e-r2max}
\end{equation}
the derivation leading to the expression for $\alpha_-$ 
in Eq.~\ref{e-alpha-pm} is no longer valid. Instead, the 
lower limit $\alpha_-$ should be set to zero.

As mentioned above, the preselection enabled by Eq.~(\ref{e-alpha-pm}) can be
applied in two ways:
\begin{list}{(\roman{enumi})}{\usecounter{enumi}}
\item a point that fails to be a member of a ``close'' pair for a given maximum
separation $r_2$ in all of the 12 directions of holonomy transformations 
to adjacent copies of the point in the covering space can be removed from
the list of uniformly selected points; and
\item when iterating through pairs of points and holonomy transformations $g_i$
($i=1,\ldots,12$), a pair of points for which at least one of the two 
points does not satisfy the condition $\alpha \le \alpha_+$ can be 
rejected without calculating the spatial separation of the pair.
\end{list}
When $r_2$ is large, $\alpha_+$ will also be large, and 
the six pairs of annuli may be sufficiently wide
that together they cover the whole sky. In this case, 
effect (i) will not occur. For a small enough separation $r_2$, the effect
should occur. In that case, more points can be uniformly selected from $S^2$
according to a uniform distribution, and again tested, until the required
number of points is obtained. In that case, a higher fraction (though not 100\%)
of pairs defined by this set of points will be useful for the cross-correlation
function calculation. This should increase the number of pairs per bin, 
especially for the smallest bins, which have the fewest numbers of pairs.
This would be useful for high resolution calculations.
Both effects (i) 
and (ii) should increase the calculation speed for a
given number of points $N_p$, since they avoid having to carry out unnecessary
matrix multiplications.

\subsection{Use of preselection on the five-year WMAP data}
\label{s-method-fiveyr}

The MCMC analysis is performed as in RBSG08, 
using the ILC and TOH five-year WMAP data (\SSS\ref{s-wmap}), but
including the smallest separation bin, i.e. using the full range of separations
\begin{equation}
d/\rSLS  < 40/90,
\label{e-range-dimless}
\end{equation}
i.e. 
\begin{equation}
d  \ltapprox  4.4 {\hGpc}
\label{e-range-Gpc}
\end{equation}
for $\Omtot = 1.01$, 
matter density $\Omm = 0.28$ and SLS redshift $z_{\mathrm{SLS}} = 1100$,
corresponding approximately to angles on the SLS
$ 0 \ltapprox \theta_d  \ltapprox 25\ddeg$.
All other 
parameters are kept as in \SSS{3.6} of RBSG08. 
\postrefereechanges{In particular, this
includes the five parameters for orientation of the fundamental dodecahedron
(galactic longitude and latitude of one face centre $(l,b)$ and a rotation
parameter $\theta$ around the axis defined by $(l,b)$), the matched circle
size $\alpha$ and the ``generalised'' twist phase $\phi$ when matching opposite
faces.}
The GPL (GNU General Public Licence) program {\sc circles}\footnote{Version
{\sc circles-0.3.2.1} was used for calculating the MCMC chains 
for the five-year WMAP maps, and version
{\sc circles-0.3.8} was used 
for calculating the chains for the INC3 observational and simulational maps.
Various versions of {\sc circles} are downloadable from 
\url{http://adjani.astro.umk.pl/GPLdownload/dodec/}. These and earlier
versions of the software 
require medium to advanced {\sc GNU/Linux}, 
{\sc Fortran77} and {\sc C} experience for a scientific user.}
is used.

\tbench

\subsection{Analysis of simulations} \label{s-method-GRF}

Although the method presented in \SSS\ref{s-method-anglimits} should
speed up the calculation, analysing a large number of simulations is
still time-consuming. 
Here, we estimate $\Sxi$ for 50 simulations and the WMAP observational map 
described in \SSS\ref{s-inc-sims} and select those 20 simulations whose 
$\Sxi$ estimates are smallest, 
provided that $\Sxi$ is not less than $\Sxi$ of the observational map.
\postrefereechanges{In each estimate, 10,000 points selected randomly 
from a uniform distribution on the sky outside of the kp2 mask
were used.}
Using each of these 20 simulations and the observational map, 
four MCMC chains with random starting points
in the parameter space described in Eq.~(28) of RBSG08 are 
carried out.

\section{Results} \label{s-results}

\subsection{Benchmarking} \label{s-res-bench}

An example set of calculation times and number of pairs in 
the smallest separation bin are shown in Table~\ref{t-bench}, for one
calculation of the
auto-correlation and cross-correlation functions at an
arbitrary PDS orientation and twist, for $\alpha=5\ddeg$
and $\alpha = 60\ddeg$, which determine the ratio $\rSLS/\rC$.
%
%
%
The speed-up factors range from about 3--10, depending on both $\alpha$ and $r_2$.

For $d < r_2 = 4.4\hGpc$, the ``annulus outer radii'' are
$\alpha_+ = 45.0\ddeg, 90.9\ddeg$ for $\alpha = 5\ddeg, 60\ddeg$ respectively.
This is clearly too large to allow any removal of points from the list of potentially
useful pairs, i.e. effect (i) in \SSS\ref{s-method-anglimits} does not
occur: $\Nsc$ and $\Nmc$ are negligibly affected by the pair
preselection mechanism. 
However, the labelling of points to record which annuli they can 
potentially be pairs of does yield a speed-up through effect (ii), i.e.
by factors of about 7 and 3 
for $\alpha = 5\ddeg$ and $  60\ddeg$ respectively.

For a much smaller maximum pair separation, i.e. $d < r_2 =
0.4\hGpc$, even though the annulus outer radii are still quite large,
i.e.  $\alpha_+ = 13.6\ddeg, 62.9\ddeg$ for $\alpha = 5\ddeg, 60\ddeg$
respectively, both effects (i) and (ii) occur. That is, not only is
there a speed-up by a factor of about 7-10, but there is also an
increase in the number of pairs in the smallest bin for calculating
the cross-correlation function, by factors of about 30 and 7
for $\alpha = 5\ddeg$ and $  60\ddeg$ respectively.

In practice, use of a small maximum pair separation bin to get
cosmologically significant results will be complicated by the relatively
larger contributions from the Doppler and ISW effects, from residual foreground
contamination, from the differences between various versions of the
all-sky CMB map at these resolutions \nocite{Aurich2005circ}({Aurich} {et~al.} 2006), and from 
the absence of the large scale signal. 
Nevertheless, the concern expressed in {\SSS}5.5 of RSBG08 that improvements
to the algorithm would be geometrically quite complex, at the risk of 
introducing biases to the method, appears to have been overcome, reducing
one obstacle to small scale work using this type of method.

In a full MCMC chain, the matched circle size $\alpha$ will vary
between the limits illustrated in Table~\ref{t-bench}. The actual
speed-up factor (and increase in numbers of pairs per bin if this
occurs) will depend on the particular path of the MCMC chain through
parameter space.
In the calculations leading to the results described below, the
range $5\ddeg < \alpha < 60\ddeg$ was retained.

\ffivelbthN

\tdodec 

\falphaphi

\talphaphifiveyear

\fphihist

\subsection{Parameter estimates from the five-year WMAP data} 
\label{s-res-lbtheta}

For both maps of the five-year data as described in \SSS\ref{s-wmap} (ILC and TOH),
$\Nchain=\Nchainsmain$ 
MCMC chains were run, each starting with different random seeds,
using the kp2 mask. Each run had 12,000 steps and the first 2000 steps of each
were discarded.\footnote{The MCMC chains used in this paper can be downloaded for 
independent analysis from the file
\protect\url{http://adjani.astro.umk.pl}
\protect\url{/GPLdownload/MCMC/mcmc_RBG08.tbz}.}
Fig.~\ref{f-five_lbth_N} 
shows the sky positions $(l,b)_{i=1,12}$ implied
by the $(l,b,\theta)$ triples in the MCMC chains 
for which $P > 0.5$ [Eq.~(25), RBSG08], for the ILC and TOH
versions of the five-year WMAP observational data. 
\postrefereechanges{The ``probability'' 
function $P$ used for optimisation by the MCMC procedure is that defined
in Eq.~(25) in RBSG08. This is not a true probability function.}


Similarly to what was done in RBSG08, the {\Nchainsmain}
chains are grouped together into four
groups, each of {\Npergroup} chains.
For a given group, steps 2001 to 12,000 from each of the {\Npergroup} chains are
concatenated. 

We make the convergence requirements a little more stringent than was
described in Sect.~4.1 of RBSG08. 
That is, 
we start from an initial angular radius of $\beta_1 = 30\ddeg$ (covering
most of the sphere), decrease by $1\ddeg$ for the next 20 iterations, 
and then remain constant at $\beta_{j\ge 21} = 10\ddeg$ until the iteration
for that face number converges or until a total of 40 iterations has been
reached.
The analyses of the 
four concatenated groups of chains give what are considered to be four independent
estimates in order to get an estimate of the uncertainties due to our
MCMC estimation method. 

The resulting numerical estimates are listed in Table~\ref{t-dodec}.
The columns show
minimum ``probability'' $\Pmin$,
face number $i$, 
number $n$ of MCMC steps contributing to the estimate obtained
from the final iteration,
galactic longitude $\lII$ and latitude $\bII$, 
and
the standard error in the mean between the four estimates of different
sets of MCMC chains, in great circle degrees, $\sigmalbth$. 
These values do not differ significantly from those in 
Table~1 and Table~4 (for the kp2 mask) of RBSG08.

The MCMC states for $\alpha$ and $\phi$ in the final radii of convergence,
and their means and standard errors in the mean, are shown in 
Fig.~\ref{f-alpha-phi} and Table~\ref{t-alpha-phi-5yr}.
\postrefereechanges{Histograms of the distribution of $\phi$ 
are shown in Fig.~\ref{f-phi_hist}.}
Again, these values do not differ significantly from those in 
Table~2 of RBSG08. 
\postrefereechanges{The TOH map shows a small offset between the 
best estimate of $\phi$ found 
(for $P > 0.5$) 
using our convergence algorithm and that at which the histogram of
$\phi$ states peaks. The former is $\phi = 30.4\ddeg$, i.e. a little
below $36\ddeg$, while the latter (lower panel of
Fig.~\ref{f-phi_hist}) is a few degrees above $36\ddeg$. The
difference can reasonably be attributed to the fact that the former
uses a convergence algorithm in multi-dimensional parameter space,
whereas the latter is the peak of a projected distribution using a
single parameter.}

\tSxifiveyr

As was discussed in Sect.~4.2 of RBSG08, when $\alpha$ is not too large,
small changes in the comoving separation between pairs 
lead to relatively large changes in $\alpha$. Does the use of the
smallest bin in the present analysis, and/or the use of the slightly
improved quality between the three-year and five-year WMAP data help
overcome the large uncertainty in $\alpha$?  
Visual inspection of Fig.~\ref{f-alpha-phi} suggests that the MCMC chains
on the ILC five-year map favour
$22 \ddeg \ltapprox \alpha \ltapprox 35\ddeg$, while those on the TOH map
favour
$15 \ddeg \ltapprox \alpha \ltapprox 35\ddeg$. 
The problem of approximate degeneracy in $\alpha$, which is
presumably sensitive to moderate levels of systematic error, remains.

\subsection{Simulations}
\label{s-res-sims}

\subsubsection{Estimates of $\Sxi$}  \label{s-res-Sxi}

The value of $\Sxi$ [Eq.~(\ref{e-defn-Sxi})] in the INC3 observational map 
\postrefereechanges{(using the kp2 mask)} is
\begin{equation}
\SxiINCthr = 963 (\mu K)^4.
\label{e-inc3-Sxi}
\end{equation}
The 50 simulations have $\Sxi$ in the range 
$1170  (\mu K)^4 < \Sxi < 8645 (\mu K)^4$, i.e. up to about an order of magnitude
higher $\Sxi$ than in the observations. As mentioned in 
\SSS\ref{s-inc-sims}, this is because 
the WMAP cosmological signal has a lot of power
close to the Galactic Plane and we use the kp2 cut sky. 
In order to use
the simulations best able to test the hypothesis that a simply connected
universe with an observationally valid large scale auto-correlation can
give a PDS-like signal, we select the 20 of these with the lowest $\Sxi$
values, i.e. in the range 
\begin{equation}
1170  (\mu K)^4 < \Sxi < 3782 (\mu K)^4,
\label{e-sims-Sxi}
\end{equation}
i.e. with up to about 3.9 times larger $\Sxi$ than in the
observations.  Since at large scales (outside of the kp2 cut), these
simulations are more correlated than the observations, we should
obtain an upper limit to the the frequency of detecting PDS-like
signals, assuming that the PDS hypothesis is wrong.

As a check on the amount of large scale power present in the Galactic Plane
in different versions of the map of cosmological signal, we estimate
$\Sxi$ for the ILC and TOH five-year maps with and without the kp2 mask.
Table~\ref{t-Sxi-fiveyr} shows these estimates. Both maps have much more power
without the cut than with the cut, as expected. Moreover, while the difference
in estimates of $\Sxi$ for the two maps differs by nearly a factor of two 
for the full sky, it differs by only 10\% for the cut sky. This confirms the
advantage in analysing the cut sky rather than the full sky: there is approximate
consensus between these two different methods of generating the map.
The INC3 estimate of $\Sxi = 963 (\mu K)^4$ is also close to these two estimates.

\subsubsection{MCMC chains}  \label{s-res-MCMC-analysis}

While the speed improvement introduced in this paper makes it possible
to analyse a set of simulations rather than just an observational map, 
carrying out large numbers of MCMC chains still remains computationally
prohibitive.
For this reason, 
we carry out only a small number of MCMC chains (four) on each data set
(INC3 observational data set or simulation), and do not attempt to estimate
uncertainties on the individual optimal parameters for a given data set.

\fINCthreelbthN

\fINCthreelbthNsim

\talphaphiINCthr

We first concatenate together steps 2001 to 12,000 of each chain in a group,
i.e. we ignore 2000 burn-in steps.
This concatenated chain is considered to be a single chain 
for the iterative procedure of estimating parameter values.
We modify 
the method of choosing an initial rough estimate of
the optimal dodecahedral face centre positions that is used to start the
iterations
towards a more precise estimate by randomly
selecting a set of dodecahedral face centres [$(l,b)$ chosen from a
uniform distribution on $S^2$, $\theta$ chosen from a uniform
distribution on $(0,2\pi/5)$.] This risks causing the convergence to 
be less accurate, but since this is applied in the same way for both
simulations and observations, this should not introduce any statistical 
bias for comparison of observations to simulations.

We also make our iteration a little more stringent than that 
described in \SSS\ref{s-res-lbtheta}. In the first iteration of 
parameter estimation, we estimate the dodecahedral face centres starting
from a random initial set as just described, and in the following 
iterations, we converge on both dodecahedral faces centres and $\alpha$,
$\phi$ simultaneously. The values of $\beta_j$ (see \SSS\ref{s-res-lbtheta})
are unchanged.  The set of points within the convergence radius 
$\beta_{j \ge 21} = 10 \ddeg$ 
is used for the final estimates of $\alpha$ and $\phi$.

\falphaphiINCthree

\falphaphiINCthreesim

\subsubsection{Optimal dodecahedron orientation: $(l,b,\theta)$ space}

Figure~\ref{f-INCthree_lbth_N} shows the optimal set of face centres
resulting from the four MCMC chains for the INC3 observational map.
While the sharpness of the optimal signal does not appear as strongly as
in the results using larger numbers of chains in RBSG08 and in 
Fig.~\ref{f-five_lbth_N} for the five-year data here, it is clearly 
consistent in position. 
Since we use many fewer chains here, it is unsurprising that the
signal appears weaker. This should not be a problem form the
present purpose, since the uncertainties from using a small
number of chains should be statistically equivalent for both 
INC3 observational data and simulations.

\falphaphicfsims

\tphisims

On the other hand, the relative weakness of the signal and the smaller
numbers of chains require using slightly lower minimum probability
thresholds for estimating the best solution. Below, we cite results
for $\Pmin = 0.3, 0.4, 0.5$ rather than the earlier thresholds of
$\Pmin = 0.4, 0.5, 0.6$.

Examples of optimal dodecahedral orientations, for the two simulations
whose $\Sxi$ estimates are lowest (i.e. are closest to that of the 
observations), are shown in Fig.~\ref{f-INCthree_lbth_N_sim}. In one case,
there is clearly a problem in converging on a single set of dodecahedral
face centres, while in the other, there does appear to be more or less 
convergence to a single solution.

\falphaSsims

\fINCthreelbthNsimhiS

\falphaphiINCthreesimhiS

\subsubsection{Circle size $\alpha$ and twist phase $\phi$} 
\label{s-res-sim-alpha}

Figure~\ref{f-alpha-phi-inc3} shows that convergence in $\alpha$ and
$\phi$ occurs for the INC3 observational map in a similar way to that
of the five-year observational maps.  In contrast,
Fig.~\ref{f-alpha-phi-inc3_sim} shows that the two simulations whose
$\Sxi$ estimates best match that of the data, i.e. those whose best
dodecahedral face centres are shown in
Fig.~\ref{f-INCthree_lbth_N_sim}, both ``escape'' towards the lower
limit $\alpha = 5\ddeg$.  The median circle size is $\alpha = 5\ddeg$
in both cases.  This effect can be expected due to the increased
relative Poisson error when comparing fewer numbers of pairs of
pixels. \nocite{Aurich08a}{Aurich} (2008) found a similar effect, describing it as
``drifting'' towards ``large $L$'', which corresponds here to small
$\alpha$.  Several previous authors have found that independently of
whether or not a cosmic topology signal is present, relatively higher
cross-correlations for zero separation pairs, i.e. for pairs on
exactly matched circles, have been found to occur as the circle size
$\alpha$ approaches zero.  For example, see Fig.~2 in \nocite{CSSK03}{Cornish} {et~al.} (2004),
Figs~4--6 in \nocite{RLCMB04}{Roukema} {et~al.} (2004), or Fig.~3 in \nocite{LR08}{Lew} \& {Roukema} (2008), where various
definitions of a normalised correlation statistic $S$ are shown to
increase as $\alpha$ decreases towards zero. Hence, it can be expected
that MCMC chains will be drawn towards the lower $\alpha$ limit.

\fphihistsims

In cases where this occurs, chains spend a large amount of time at
this limit, but cannot go below it.
For this reason, the 
median of the $\alpha$ 
estimates (above a minimum ``probability'' used in the MCMC chains,
e.g. $P > \Pmin = 0.5$), should more accurately represent the 
optimal region favoured by the MCMC chains than the mean. Thus, here
we use the median of $\alpha$. 

Figure~\ref{f-alpha_phi_cf_sims} shows that nearly all the simulations escape to
the lower $\alpha$ limit. Most of the simulations have
best estimate $\alpha$ values less than $10\ddeg$, and two have
$\alpha$ estimates just a few degrees higher. Only two of the simulations have 
an optimum $\alpha$ estimate anywhere near the circle size
$\alpha$ of the INC3 observational best estimate given in 
Table~\ref{t-alpha-phi-INC3}: $\alphaINCthr = 30.8\ddeg$ (for $\Pmin=0.5$).

\postrefereechanges{Do} the latter two 
simulations, which are similar to the observational map
in the sense that an optimal solution away from 
low $\alpha$ limit is found, have sufficiently similar characteristics to 
the observational map such that the observational map can be considered
a chance realisation statistically similar to these two simulations?
Figure~\ref{f-alpha_S_sims} does not support this.
The two simulations are among the simulations that have the highest $\Sxi$
estimates, well above three times that present in the observational
map. This is consistent with what is expected: the stronger the
auto-correlation on large scales, the higher the chance is of getting
high cross-correlations between apparently distant parts of the sky over
a large number of pixel pairs, rather than escaping to the low $\alpha$
limit where there are relatively few pixel pairs.
This can be quantified as follows.

Since the $\alpha$ distributions are constrained from below, they are
unlikely to be Gaussian. So, estimating the significance of the
correlation between $\alpha$ and $\Sxi$ among the simulations 
is best done using a non-parametric
statistic. Spearman's rank correlation $\rho$ and Kendall's rank correlation 
$\tau$ one-sided tests with a positive correlation as the alternative hypothesis
give probabilities that $\alpha$ and $\Sxi$ are unrelated of 2.5\% and
2.4\% respectively.  Even if we
arbitrarily remove the two simulations
with high $\alpha$ from the data set, the same two two rank correlation tests
give probabilities that  $\alpha$ and $\Sxi$ are unrelated of 
12.6\% and 12.5\% respectively. 

This supports the visual impression from Fig.~\ref{f-alpha_S_sims}.
For low $\Sxi$, an optimal cross-correlation
in GRF simulations should escape to the low $\alpha$ limit.
The point representing the observational simulation appears to be quite
exceptional. 


Do the two simulations with high $\alpha$ estimates (numbered 58 and
80, with $\alpha$ estimates of $24.5\ddeg$ and $31.9\ddeg$
respectively) have convergent estimates of dodecahedral face centres
and $\phi$?  Figure~\ref{f-INCthree_lbth_N_sim_hiS} indicates a poor
convergence 
of dodecahedral face centres
for simulation 58 and what looks like the superimposition of 
a strong primary and a weak secondary convergence in the case of 
simulation 80. Moreover, Fig.~\ref{f-alpha-phi-inc3_sim_hiS} shows that 
both of these two simulations have quite strongly 
bimodal distributions in $\phi$ rather than favouring any individual optimal
value of the twist $\phi$. This is quite different behaviour to that
in Fig.~\ref{f-alpha-phi-inc3} for the observational map.

However, in order to be conservative, let us suppose that simulations 58 and
80 converge well enough in comparison to the observational map that we can
consider them to have convergent MCMC solutions with 
$\alpha \gg \alphalimit = 5\ddeg$.
This gives us an estimate
\begin{equation}
P( \alpha \gg \alphalimit \;|\; \Sxi \le 3.9 \SxiINCthr) \ltapprox 10\%,
\label{e-prob-PDS-given-Sxi_alpha}
\end{equation}
where $\alpha \gg \alphalimit$ represents the event of getting
a non-Poisson-noise signal at least $10\ddeg$ (the MCMC step size) 
away from the lower limit of $\alphalimit$.


\subsubsection{Distribution of the optimal twist phase $\phi$}

It is clear in Fig.~\ref{f-alpha_phi_cf_sims} that the simulations
give a distribution of best estimate twist angles $\phi$ different
from the uniform distribution on $[-\pi,\pi]$ described in Eqs~(12)
and (13) of RBSG08 and discussed in Sect.~5.4 of that paper.
\postrefereechanges{Figure~\ref{f-phi_hist_sims} shows the
  distribution as a histogram.}  A two-sided Kolomogorov-Smirnov test
between the \postrefereechanges{distribution of the best estimates of
  $\phi$ from the 20 simulations} and a uniform distribution on
$[-\pi,\pi]$ rejects equality with $P=0.01$. The values of $\phi$
\postrefereechanges{for the simulations}, shown in
Figs~\ref{f-alpha_phi_cf_sims} and \ref{f-phi_hist_sims}, all lie in
the range $(-100\ddeg,+100\ddeg)$, and mostly seem to cluster even
closer to $\phi=0$.

A likely explanation is that this is a consequence of the
anti-correlation in the auto-correlation function measured in WMAP sky
maps at nearly antipodal scales by different authors using different
methods (Fig.~16, \nocite{WMAPSpergel}{Spergel} {et~al.} 2003; Fig.~1, RBSG08).
The \nocite{Hinshaw06}{Hinshaw} {et~al.} (2007) $C_l$ values should implicitly include
the information that there is an antipodal anti-correlation.
Estimates of the auto-correlations of the simulations confirm that an
anti-correlation of about 
$-100(\mu \mathrm{K})^2$
at $\phi = \pm \pi$ is present, so that the
simulations correctly reproduce this characteristic of the
observational data.  
This anti-correlation implies that MCMC chains
should disfavour $\phi = \pm \pi$ when correlating pairs on a matched
circle pair, and, hence, generally 
\postrefereechanges{disfavour $\phi = \pm \pi$ for pairs on ``matched
annuli''.}
This is discussed further in
\SSS\ref{s-why-phi-distbn}.

Since a uniform distribution on $[-\pi,\pi]$ is clearly wrong, 
a reasonable hypothesis must be made regarding the
intrinsic, expected distribution 
\postrefereechanges{of best estimates of} $\phi$. Given the
numerical results from the simulational analyses and the presence
of the $\phi = \pm \pi$ anticorrelation, which should favour
$\phi$ away from $\pm \pi$ and towards zero,
we assume 
a Gaussian distribution, $f(\phi)$, centred on $\phi=0$ with 
width $\sigma_\phi$ estimated by the r.m.s. of $\phi$ in the simulations. 


The parameters of this distribution for different $\Pmin$ values, and
the Kolmogorov-Smirnov probabilities that the simulational values are
consistent with a Gaussian distribution, are given in
Table~\ref{t-phi-sims}.  For the lower thresholds, $\Pmin = 0.3 $ and
$\Pmin = 0.4$, the Gaussian distribution hypothesis is mildly rejected
by the Kolmogorov-Smirnov test. If we set the mean and standard
deviation of the Gaussian distribution to be the mean and standard
deviation of the simulational $\phi$ estimates, then the three
Kolmogorov-Smirnov probabilities for $\Pmin = 0.3, 0.4, 0.5$ are
$P_{\mathrm{KS}} = 54\%, 84\%, 97\%$ respectively.  Thus, it is clear
that the distributions are consistent with Gaussianity if we use the
estimated means rather than force symmetry about $\phi=0$.

Could there be a reasonable justification for using a non-zero mean $\phi$?
The only possible source of systematic asymmetry is that the noise 
simulations follow noise patterns in the observational data, which are
not perfectly symmetrical. It is not obvious that this asymmetry is 
sufficient to justify assuming an asymmetry in the expected distribution 
of $\phi$. Moreover, the means and standard errors in the mean for the 
three $\Pmin$ thresholds are 
$12.8 \pm 7.1 \ddeg$, 
$14.4 \pm 8.4 \ddeg$, 
$12.1 \pm 8.6 \ddeg$ respectively, showing no statistically significant
difference from zero.

Given a Gaussian distribution in $\phi$ centred on zero with 
width $\sigma_\phi = \phirms$ as listed in Table~\ref{t-phi-sims}, 
the probability that $\phi$ is as close to $\pm \pi/5$ 
as the observational value (Table~\ref{t-alpha-phi-INC3}) is
close to $+\pi/5$ is
$22\%, 12.8\%, 9.4\%$ for 
$\Pmin = 0.3, 0.4, 0.5$ respectively, i.e.
\begin{eqnarray}
  &&P \;\Big( \min(|\phi \pm\pi/5|) < |\phiINCthr-\pi/5|  
  \; \Big\vert \;   
  \Sxi \le 3.9 \SxiINCthr \Big) \nonumber \\
  && \ltapprox 22\%.
\label{e-prob-PDS-given-Sxi_phi}
\end{eqnarray}
Since the values $\phirms$ are themselves not too far from 
$\pi/5$, it is clearly not so improbable that $\phi \approx \pm \pi/5$,
compared to what would be expected from 
a uniform distribution on $[-\pi,\pi]$.
In other words, this suggests that using the present
method, an estimate of $\phi$ is not as good a discriminator between a
chance PDS-like signal and an intrinsic, physical signal as it would
be if the expected distribution were uniform on $[-\pi,\pi]$, i.e.
there is another {\em topological degeneracy} 
\nocite{Aurich2005a}(cf  {Aurich} {et~al.} 2005a) in CMB all-sky maps.

On the other hand, is it just a coincidence that $\phirms \sim \pi/5$?
In \SSS\ref{s-why-phi-distbn} below we discuss this question. It 
is possible that 
the empirical $C_l$ spherical harmonic spectrum, even with
randomised phases, may encode more cosmic topology information than 
might naively be expected.



\subsubsection{Probability of rejecting the simply connected, infinite, flat model } 

The analyses of these simulations indicate that the requirement for
the MCMC chains to avoid ``escaping'' to the lower limit in circle
size $\alpha$ gives a stronger constraint against the simply connected, infinite,
flat model 
than the requirement that could potentially exclude the PDS model,
i.e. the requirement that $\phi \approx \pm \pi/5$.

Let us write the PDS-like characteristics of the WMAP observational data
which we have tried to reproduce by simulations as follows. The data
\begin{list}{(\roman{enumi})}{\usecounter{enumi}}
\item have a large scale cutoff in structure statistics 
\item yield a solution with $\alpha \gg \alphalimit$ and $\phi \approx + \pi/5$
when using the MCMC method for optimising 
\postrefereechanges{the cross-correlation} $\ximc$ for the ``generalised'' PDS.
\end{list}
Rewrite these as 
\begin{list}{(\roman{enumi})}{\usecounter{enumi}}
\item $\Sxi \ltapprox \SxiINCthr$
\item $\ximc$ yields $\alpha \gg \alphalimit$ and 
  $\min(|\phi \pm\pi/5|) < |\phiINCthr-\pi/5|  \le 8.4 \ddeg$,
\end{list}
where we use the ``worst'' estimate of $\phi$ from Table~\ref{t-alpha-phi-INC3},
i.e. $\phi = 27.6\ddeg$, for $\Pmin = 0.3$.
For convenience, we write these even more compactly as 
\begin{list}{(\roman{enumi})}{\usecounter{enumi}}
\item $\Sxi \ltapprox \SxiINCthr$
\item $\PDSlikesignal$.
\end{list}

Since both  $\Sxi \approx \SxiINCthr$ and $\PDSlikesignal$ are present in
the WMAP data, the probability of these two characteristics both 
occurring in the simply connected, infinite, flat model can be written
$P[(\Sxi \approx \SxiINCthr) \cap \PDSlikesignal] $.
From the results above, in particular from
Eqs~(\ref{e-prob-PDS-given-Sxi_alpha}) and
(\ref{e-prob-PDS-given-Sxi_phi}), we can write
\begin{eqnarray}
  P(\PDSlikesignal |  \Sxi \ltapprox \SxiINCthr) & = & 
  P\Big[ \alpha \gg \alphalimit \;\mathrm{and}\; \nonumber \\
    && \min(|\phi \pm\pi/5|) < |\phiINCthr-\pi/5|   \nonumber \\ 
    && \; \Big\vert \;  \Sxi \ltapprox \SxiINCthr 
  \Big]
  \nonumber \\
  & \le & 
  P(\alpha \gg \alphalimit \; \Big\vert \;  \Sxi \ltapprox \SxiINCthr),
  \nonumber \\
  & \le & 
  P(\alpha \gg \alphalimit \; \Big\vert \;  \Sxi \ltapprox 3.9 \SxiINCthr),
  \nonumber \\
  & \ltapprox & 10\%. 
\label{e-prob-pdslikesignal}
\end{eqnarray}
Are the probabilities that $\alpha \gg \alphalimit$ and $\phi \approx \pm \pi/5$
independent of one another, so that instead of the Eq.~(\ref{e-prob-pdslikesignal}),
we can write 
\begin{eqnarray}
 && P\Big[ \alpha \gg \alphalimit \;\mathrm{and}\; \nonumber \\
&& \min(|\phi \pm\pi/5|) < |\phiINCthr-\pi/5| 
    \; \Big\vert \;  \Sxi \ltapprox \SxiINCthr 
  \Big] \nonumber \\
&&= 
  P(\alpha \gg \alphalimit \; \Big\vert \;  \Sxi \ltapprox \SxiINCthr) \times 
\nonumber \\
&& \quad P\Big[ 
 \min(|\phi \pm\pi/5|) < |\phiINCthr-\pi/5|  
    \Big\vert \;  \Sxi \ltapprox \SxiINCthr   \Big]
  \nonumber \\
&&\approx 2.2\%,
\label{e-prob-pdslikesignal-if-independent}
\end{eqnarray}
again using Eqs~(\ref{e-prob-PDS-given-Sxi_alpha}) and
(\ref{e-prob-PDS-given-Sxi_phi})?  Spearman's rank correlation $\rho$
and Kendall's rank correlation $\tau$ one-sided tests where
$\min(|\phi \pm\pi/5|)$ decreases as $\alpha$ increases as the
alternative hypothesis both give probabilities that these two
parameters are unrelated of 7.9\%.  A two-sided test gives 16\% in
both cases. While neither of these rejections of the hypothesis that
the two parameters are unrelated is highly significant, they are
strong enough rejections that it would premature to assume that the
two parameters are independent. Hence, the probability estimate in
Eq.~(\ref{e-prob-pdslikesignal-if-independent}) cannot (yet) be
assumed to be valid.

As mentioned above, \nocite{WMAPSpergel}{Spergel} {et~al.} (2003) estimated that 
$P(\Sxi \ltapprox \SxiWMAP) \sim 0.15\%$
for an infinite, flat, cosmic concordance model, 
with a fixed spectral index of density perturbations.
\nocite{EfstNoProb03b}{Efstathiou} (2004) estimated
$P(\Sxi \ltapprox \SxiWMAP) \sim 3$--$12.5\%$. The latter calculation
reconstructs unobserved structure hidden behind the galactic mask
by assuming a simply connected model. For the purposes of 
testing the simply connected model hypothesis, this is internally consistent.

Using these two estimates to write two respective estimates of
$P(\Sxi \ltapprox \SxiINCthr) \approx P(\Sxi \ltapprox \SxiWMAP),$
we have 
\begin{eqnarray}
  && P[(\Sxi \ltapprox \SxiINCthr) \cap \PDSlikesignal ] = 
\nonumber \\
  && \quad\quad\quad\quad P(\Sxi \ltapprox \SxiINCthr) \; 
  P(\PDSlikesignal | \Sxi \ltapprox \SxiINCthr)
\label{e-defn-cond-prob}
\end{eqnarray}
by the definition of conditional probability. Hence,
from Eq.~(\ref{e-prob-pdslikesignal}),
we have 
\begin{eqnarray}
P[(\Sxi \ltapprox \SxiINCthr) \cap \PDSlikesignal ]
    &<& 0.015\%,
\nonumber \\ 
P[(\Sxi \ltapprox \SxiINCthr) \cap \PDSlikesignal ]
    &<& 1.25\%
\label{e-prob-final}
\end{eqnarray}
for the lower and high estimates of
$P(\Sxi \ltapprox \SxiINCthr)$ respectively.

In other words, the simultaneous existence of both of these two
properties of the WMAP data, one a generic characteristic of small
universe models and the other highly specific to the PDS, is unlikely
with a probability of about 99.99\% or 99\% depending on whether
the \nocite{WMAPSpergel}{Spergel} {et~al.} (2003) or \nocite{EfstNoProb03b}{Efstathiou} (2004) estimates of $P(\Sxi
\ltapprox \SxiWMAP)$ are used. This result only requires a frequentist
approach to probabilities. Bayesian modelling of prior beliefs is not
invoked.

\section{Discussion}
\label{s-disc}

\subsection{Matched annuli and calculational speedup} \label{s-disc-speed}

The preselection method described in \SSS\ref{s-method-anglimits} 
leads to faster calculation times by a factor of about 3--10
(\SSS\ref{s-res-bench}). This is by eliminating most calculations of
pair separations for pairs which are not useful for the cross-correlation 
calculation. Is it possible to improve this algorithm even further?
Given that we have Eq.~(\ref{e-alpha-pm}), the number of calculations 
required for a given total number of pairs could, in principle, be 
reduced by another small factor as follows.

Randomly select a point $p_i$ from a uniform distribution on $S^2$,
considered to be the left-hand copy of the SLS in either Fig.~\ref{f-alpha_p} or
Fig.~\ref{f-alpha_m}.
For each of the 12 holonomy transformations $g_j$ to adjacent copies
of the fundamental domain, check if the angle $\alpha_{ij}$ on the SLS
from $p_i$ to the dodecahedral face centre for $g_j$ satisfies 
$\alpha_- \le \alpha_{ij} \le \alpha_+$. If this constraint is satisfied,
then choose a second point $p_i'$ randomly from a uniform distribution 
on the circle defined by the intersection of the right-hand copy of the SLS
in Fig.~\ref{f-alpha_p} or Fig.~\ref{f-alpha_m} and 
the 2-sphere centred at $p_i$, having radius $r_2$.

By construction, all of the pairs $(p_i, p_i')$ selected in this way are
at separation $r_2$, and should be statistically equivalent to generating
a full set of pairs of which both members are uniformly selected on $S^2$
and then selecting those whose separations are close to $r_2$. A loop over
values of $r_2$ will give cross-correlations over the desired range of
separations. 
This construction would bring the calculation method closer to the 
identified circles method itself, with the difference that instead
of correlating points lying precisely along the circles, points ``near'' the
circles are correlated. 

This raises the question of differences
between using the present method and using the identified circles method
with ``thickened'' circles. The most likely differences between these
two methods would depend on how ``thickening'' is defined and how
overlapping thickened circles are dealt with. 

``Thickening'' would require a somewhat arbitrary choice of an
averaging procedure. For example, a circle could be divided into equal
angular intervals and thickened so that individual bands of the
annulus are internally averaged to obtain the temperature fluctuation
at that angular position around the circle/annulus. Alternatively, 
a Gaussian smoothing could be used. However, neither of these methods
would take into account the fact that the changes in the spatial 
geodesic pair separation
of points inside and outside the zero thickness circle are not linear
with change in their angular distance from the centre of a circle.
In contrast, the present method calculates and uses the spatial separations
directly.

The problem of overlapping thickened circles follows from the fact
that twelve pairs of annuli on the 2-sphere intersect each other many
times.  The greater the thickness of the annulus, the greater the
number of points in these intersections.  If a correlation statistic
is calculated over each circle separately and then averaged, then
points which are members of these intersections contribute several
times (typically two or three) more to the final statistic than points
which are members of only one annulus. If chance fluctuations at some
pairs of these intersections happen to have high correlations or
anti-correlations, then the final statistic might be biased by these
pairs, since it would implicitly assume that the pairs are independent
of one another, even though this is false in some cases.

The use of numerical simulations should make this more of 
a problem of excessive noise rather than systematic bias. It is 
difficult to see any simple way in which a method directly based
on the identified circles principle could avoid this problem, whether or
not its final effect is statistical bias or rather an extra source of noise.
For the SLS optimal cross-correlation method, this problem does not exist,
since both points are selected uniformly on $S^2$ and selected afterwards
based on their spatial separations (the method as presented in RBSG08),
or else preselected in a way that is equivalent to this (the present paper).

Hence, results from comparing a series of thickened
circle pairs for a range of thicknesses to use of the SLS optimal
cross-correlation method should be expected to be different, for both
geometrical and statistical reasons.

\subsection{Five-year WMAP data} \label{s-disc-fiveyr}

The five-year WMAP data (\SSS\ref{s-res-lbtheta}) give best estimates
of the PDS model similar to those for the three-year data. Since most
of the improvement in the WMAP data is at small angular scales, 
this is unsurprising.

\subsection{Simulations} \label{s-disc-sims}

Given the infinite, flat model with GRF as a null
hypothesis, 
the probability that the observed WMAP data could be a random realisation
of this model, i.e. that both a large-scale cutoff in power and a specifically
PDS-like signal appear in the WMAP data, is estimated in 
Eq.~(\ref{e-prob-final}) as 0.015\% or 1.25\% depending respectively 
on whether we use the \nocite{WMAPSpergel}{Spergel} {et~al.} (2003) or \nocite{EfstNoProb03b}{Efstathiou} (2004) estimates
of the former.
This appears to be a significant rejection of the infinite flat model.
Is there any way to avoid this interpretation?
Other properties of the cosmological component of the WMAP data,
unlikely in the infinite flat model, have been noted by many authors.
It is unlikely that these different properties are statistically
fully independent of one another.

On the other hand, in this case we have a physically motivated
model, motivated from the most fundamental level: the spatial
section of the Universe is a 3-manifold, i.e. it must have a ``shape''. 
As \nocite{Schw00,Schw98}{Schwarzschild} (1900, 1998) stated a little over a century
ago, that shape may well be multiply connected rather than simply connected.

More recently, the generic predictions of a cutoff in power at large
scales were made when only the COBE data were available 
\nocite{Star93,Stevens93}({Starobinsky} 1993; {Stevens} {et~al.} 1993). The identified circles principle
\nocite{Corn96,Corn98b}({Cornish} {et~al.} 1996, 1998), of which the present method is an extension, 
and the matching of the Poincar\'e Dodecahedral Space hypothesis 
to the large scale lack of power and the estimates of 
$\Omtot \sim 1.01$--$1.02$ \nocite{LumNat03}({Luminet} {et~al.} 2003), were published without
the knowledge that the present version of the method, allowing
for a ``generalised'' PDS of arbitrary twist angle, would be
applied to the WMAP data.
It seems physically unreasonable, therefore, not to combine the probabilities
of the different signs of cosmic topology.

In this case, to argue that the probability in Eq.~\ref{e-prob-final} has
underestimated, either the 
\nocite{WMAPSpergel}{Spergel} {et~al.} (2003) or \nocite{EfstNoProb03b}{Efstathiou} (2004) probability estimates of a low
$\Sxi$ have
to be increased, or the estimate of the probability of a PDS-like signal,
given a low $\Sxi$, (Eq.~\ref{e-prob-pdslikesignal}) has to be increased.
We have conservatively taken the maximum probability estimate 
in Table~5 of \nocite{EfstNoProb03b}{Efstathiou} (2004) for the former. 

For the latter, we have conservatively assumed that the two
simulations with $\alpha \gg \alphalimit$ have a signal given by the
MCMC method similar enough to the signal in the observational map that
we can set this probability at 
$  P(\PDSlikesignal |  \Sxi \ltapprox \SxiINCthr) \le 10\%$.  To increase this
probability significantly, it would be necessary to argue that {\em
  many or most} of the simulations have convergence characteristics,
best estimates of $\alpha \gg \alphalimit$ and best estimates of $\phi
\approx \pm \pi/5$ similar to those for the observational map. 
Given the results presented in \SSS\ref{s-res-sims}, a systematic error
of this sort able to satisfy this seems unlikely.

\fsymsig

On the contrary, Fig.~\ref{f-alpha_S_sims} and the 
rank correlation statistics (\SSS\ref{s-res-sim-alpha}) favouring 
a positive correlation between $\alpha$ and $\Sxi$ suggest that 
$  P(\PDSlikesignal |  \Sxi \ltapprox \SxiINCthr)$ is {\em smaller} than what
we have been able to estimate with a small number of simulations.
This is consistent with what was
argued in RBSG08, i.e. that
the lower the amplitude of the large-scale
auto-correlations, the less chance there should be of cross-correlations
occurring in the absence of a PDS-like signal or for a wrong orientation
of the PDS model.

This suggests two possible alternative approaches to that used here.
Either we could generate simulations so that {\em most} of the set
of simulations have $\Sxi$ about as low as that observed --- with the kp2 cut ---
or we could generate simulations for the full uncut sky. However, both
of these approaches have problems.
The problems in both cases arise from the fact that a large amount of
the large scale power is estimated to lie close to the Galactic Plane, 
but the estimates of how much this power is and of how precisely we can
measure the detailed fluctuations close to the Galactic Plane are highly
uncertain. For example, Table~\ref{t-Sxi-fiveyr} shows that the estimates
for $\Sxi$ for the five-year ILC and TOH maps differ by only 10\% when
the kp2 cut is used, but differ by nearly a factor of 2 for the full sky.

Suppose that we generate simulations so that most of the set
of simulations have $\Sxi$ about as low as that observed, using the kp2 cut.
In this case, 
we implicitly assume that both the extrapolations from outside of the kp2 cut
to inside it, as well as the direct estimates for the full sky, vastly
overestimate the power inside the kp2 cut. For testing a multiply connected
model hypothesis, this could have some validity given that the spherical harmonics
are not statistically independent from one another. However, this is not
the null hypothesis that is to be tested with the simulations. The aim is
to test the null hypothesis of a simply connected, infinite, flat model
with Gaussian random fluctuations. This model implies a relation
between structures inside and outside of the kp2 cut, which needs to be
included in the simulations if we wish to correctly test the hypothesis.
In other words, the statistics of
fluctuations generated by simulations which are designed to 
mostly have small $\Sxi$ 
outside of the kp2 cut are unlikely to be statistically equivalent to 
those of fluctuations generated in the way performed here. 
In the present work, 
the full set of simulations was generated 
by the \nocite{Hinshaw06}{Hinshaw} {et~al.} (2007) estimates of the $C_l$ values, and a subset of 
these was selected with the criterion that $\Sxi$ outside of the kp2 cut 
must be as close as possible to $\Sxi$ of the observations outside of the kp2 cut.
These two different methods of generating simulations are distinct.

The second alternative to the present method would be to generate
simulations for the full uncut sky, and to run MCMC chains on both
these and the observational map. However, because of the intrinsic
difficulty in correcting for emission in the Galactic Plane and as is
indicated in Table~\ref{t-Sxi-fiveyr} for two different versions of
the all-sky map of the cosmological signal, there would then be a much
greater systematic uncertainty in the results.
Hence, both of these two alternative approaches have disadvantages relative
to the method used here.

\fphihistINC

In addition to our main results of probability estimates, it has become
clear that the properties of the expected distribution of $\phi$ (for the
null hypothesis of an infinite, flat model) are not simple. This can be
understood generically by realising that the mathematical procedure 
we are using is pattern matching. A cross-correlation for the correct 
mapping between two genuinely correlated copies of a single pattern will
necessarily yield a high value. However, for a pattern which is sufficiently
complex, both cross-correlations for incorrect mappings between two genuinely
correlated copies of a single pattern, and also cross-correlations for 
arbitrary mappings between two uncorrelated patterns may in some cases
yield high values due to the complexities of the patterns and chance
correlations. From first principles, modelling this is unlikely to be
simple. For this reason, simulations provide an algorithmic shortcut
to estimating the likely distribution of $\phi$, given a certain family
of patterns. Gaussian random fluctuations from a given $C_l$ spectrum
are one such family of patterns. For a different $C_l$ spectrum or
different statistical properties of the fluctuations, different 
characteristics of the expected distribution of $\phi$ may occur.
Here we discuss some characteristics of interest.

\subsubsection{Tendency of an MCMC chain to favour regions of $\phi$ 
approximately symmetric around zero} \label{s-why-phi-symmetric}

Figure~8 in RBSG08 and Figs~\ref{f-alpha-phi} and
\ref{f-alpha-phi-inc3} each indicate that in addition to the main
cross-correlation signal, there is also an additional, weak, secondary
signal with an approximately equal but opposite value of $\phi$. This
is clear in Table~3 of RSBG08 \postrefereechanges{and
  Fig.~\ref{f-phi_hist} for the five-year ILC and TOH maps.
  Since only four MCMC chains were carried out for each of the INC3
  real and simulational maps for the main INC3 analysis, we carried
  out 12 additional MCMC chains on the real INC3 map in order to create
  a histogram equivalent to those in  Fig.~\ref{f-phi_hist}. 
  The resulting histogram of the
  $\phi$ distribution from the 16 chains is shown in
  Fig.~\ref{f-phi_hist_INC3}. This is clearly consistent with those in
  Fig.~\ref{f-phi_hist}, in the sense that the $+\pi/5$ peak strongly
  dominates over the $-\pi/5$ peak.

  Is it physically reasonable that a PDS model would give both
  $-\pi/5$ and $+\pi/5$ as valid twist angles, even where one twist
  gives a much weaker signal than the other? At most one of these can
  indicate the correct 3-manifold of comoving space. A possible
  explanation for both twist angles to be favoured in the MCMC chains
  might be that the density perturbations went through some sort of
  resonance process during an early epoch. In that case, it is
  conceivable that a harmonic created at that
  early epoch would still remain present.

The} optimal cross-correlations in the
simulations, in particular those shown in
Figs~\ref{f-alpha-phi-inc3_sim} 
\postrefereechanges{and \ref{f-alpha-phi-inc3_sim_hiS}, 
also have bimodal distributions of preferred values of $\phi$.
Moreover, they}
appear to give nearly equal weight to
solutions with positive and negative values of $\phi$ of about the
same absolute value.  
\postrefereechanges{This suggests that some type of pattern
in the fluctuations can tend to cause
an MCMC chain to favour regions of $\phi$ 
approximately symmetric around zero.}

The following schematic diagram shows that at least one specific
pattern of fluctuations on the sky can lead to some degree of symmetry
of optimal $\phi$ values, to the extent to which the real (or simulated) pattern
mimics the idealised pattern.
Figure~\ref{f-symsig} shows two circles on opposite sides of the sky, 
seen by someone external to the SLS, looking approximately but not nearly
along the axis joining the two circles. If we approximate this region of the
covering space $S^3$ as an approximately flat region for intuitive 
simplicity, then the mapping corresponding to a ``generalised'' holonomy 
transformation from one circle to the other is a translation followed by 
a twist. If the twist is either
$+\phi$ or $-\phi$, then the summed
cross-correlation is 1. 
If the twist is anything other than $\pm \phi$, including zero, 
then these two idealised circles will have a zero cross-correlation. 

\fantipode

The original identified circles principle, i.e. the present method in 
the limit of zero pair separations, would imply that both of these patterns
are equally optimal. By extension, our present method will find the
same result for this idealised pattern. 

This schematic situation is clearly highly simplified. Apart from the
fact that the fluctuation patterns are unlikely to be as simple as in
Fig.~\ref{f-symsig}, here we have assumed that the MCMC chain is fixed
at a particular orientation on the sky and matched circle size. In
reality, the chains are free to change both of these. Changing orientation
and/or circle size would weaken the symmetry provided by the pattern 
in Fig.~\ref{f-symsig}.
Hence, to
calculate the relevance of this schematic pattern realistically from
first principles would be quite complex. However, given that we have
realistic numerical simulations (for the infinite, flat model), we do
have an illustration here of at least one 
basic pattern which could
lead to 
best solutions 
which are approximately symmetric in $\phi$ around 
two values $\pm \phi$, even though only one of these two values is the 
``true'' value (physical or simulated) of the map.

\subsubsection{Properties of the distribution of optimal $\phi$ values}
\label{s-why-phi-distbn}

Statistical properties of the distribution of optimal $\phi$ values
found in the analyses of the simulations 
(Fig.~\ref{f-phi_hist_sims})
are different from what was
intuitively expected.
Not only is the distribution of $\phi$
{\em not} uniform on $[-\pi,\pi]$, but it is moderately 
well fit by a Gaussian centred on 
$\phi=0\ddeg$ of width
$\sigma_{\phi} 
\equiv \sqrt{\left<\phi^2\right>} \approx \pi/5$
(Table~\ref{t-phi-sims}). 
Why is the distribution of optimal $\phi$ values 
not uniform on $[-\pi,\pi]$? 
Why should $\sigma_\phi$ be approximately $\pi/5$? 

\falphaphiantip

\fphihistantip

Consider the twist angle $\phi=\pi$. On matched circles, this represents
pairs of antipodal points, independently of the matched circle size
$\alpha$. Figure~\ref{f-antipode} illustrates this schematically.
In general, a pair of points on a pair of would-be matching circles 
are separated by a twist $\phi \ll \gamma$, where $\gamma$ is 
the angle on the SLS separating the pair of points. The exception 
is that $\phi = \pi$ when $\gamma = \pi$. This separation corresponds
to a spatial geodesic separation between the pair of points 
in the covering space $S^3$ of $2\rSLS$.

For antipodal points, $\xisc$ is slightly negative 
(Fig.~16, \nocite{WMAPSpergel}{Spergel} {et~al.} 2003; Fig.~1, RBSG08). In the simulations
analysed here, this anti-correlation at $\phi=\pi$ is
approximately $-100(\mu \mathrm{K})^2$ in most simulations.

The MCMC chains are optimised to find positive cross-correlations, 
not anti-correlations. Hence, it should, in fact, be 
expected\footnote{This fact was missed in RBSG08.} that optimal
cross-correlations should cluster 
\postrefereechanges{towards} $\phi = 0$ and away from $\phi = \pi$.

We test this with a toy model as follows. The angular scale above
which the auto-correlation becomes negative in Fig.~16 of
\nocite{WMAPSpergel}{Spergel} {et~al.} (2003) is $\mathrm{acos}(-0.9) = 154\ddeg.$ To first
order, we can approximately think of this as antipodal pairs of
25{\ddeg} discs on the sky being anti-correlated. So, we start with one
of the simulated maps and randomly select 50 antipodal pairs from a 
uniform distribution on the
sky. For each antipodal pair, we calculate the mean temperature
fluctuation within 10{\ddeg} of each pole. We choose this radius to be
smaller than 25{\ddeg} so that there is a fair chance that the 
disc we use will mostly cover a single ``anti-correlated'' disc, of
full size 25\ddeg. If we had chosen the full disc size, then frequently there
would \postrefereechanges{be too} little overlap for an anti-correlation to be measured.
If the product of the mean temperature fluctuations in the two discs of
the pair is negative, then we multiply the fluctuations in one member of
the pair by -1. This is done for all 50 antipodal disc pairs. If there
were no overlap between these discs, then this would cover about 80\%
of the full sky. 

While this procedure is {\em not} likely to give a valid statistical set 
of fluctuation patterns for testing the infinite, flat universe hypothesis,
it should be sufficient for qualitatively 
testing the hypothesis that the anti-correlation 
is responsible for the concentration of the 
expected distribution of $\phi$ away from $\pi$ and towards zero. 
Moreover, since the anti-correlation at the antipodes provides some 
contribution to $\Sxi$, this procedure
decreases the minimum, mean, and median values of $\Sxi$ for
the 20 simulations from 
$1170, 2928, 3071 (\mu K)^4$ to 
$467, 1494, 1169 (\mu K)^4$
respectively, i.e. by about a factor of two.
The maximum value of $\Sxi$ increases slightly, from 
$3782$ to $4012 (\mu K)^4$.

Figure~\ref{f-alpha_phi_antip} shows the result of searching for
optimal PDS solutions using four MCMC chains per modified simulation,
and analysing the chains as before. The distribution of best $\phi$
estimates is shown in Fig.~\ref{f-phi_antip_hist}.
Several effects of the antipodal anti-correlation reversal
are visible in the two figures.
Firstly, several optimal correlations occur 
close to the antipodes, i.e. at $\phi \sim \pi$.
Secondly, the chains which escape to $\alphalimit$ have optimal $\phi$
values spread over a wider range than without the modification.
These two effects are qualitatively consistent with the hypothesis
that the antipodal anti-correlation plays an important role in 
concentrating the expected $\phi$ distribution for an infinite,
flat model towards zero.

A third effect can also be seen in Fig.~\ref{f-alpha_phi_antip};
a fairly large number of chains no longer escape to 
$\alphalimit$. Since we reverse the fluctuation sign in one element of
each of 50 antipodal pairs, it is clear that this does not only affect
antipodal correlations, it can also create correlations at several
different scales, which did not exist in the original simulation.
This can create highly artifical attractors for the MCMC chains.



The combination of these effects does not yield a distribution of $\phi$
which looks uniform by inspection of Figs~\ref{f-alpha_phi_antip} 
\postrefereechanges{and \ref{f-phi_antip_hist}.}
A Kolmogorov-Smirnov two-sided test comparing the $\phi$ distribution to
a uniform distribution on $[-\pi,\pi]$ yields a probability of 
16\%, i.e. 
the distribution is only marginally consistent with 
a uniform distribution. Clearly, 
although the non-uniform distribution of $\phi$ is clearly influenced
by the antipodal anti-correlation, the latter alone is insufficient 
to explain it.


Given that the observational map does include an antipodal anti-correlation,
and possibly other features contributing to the non-uniform expected 
distribution of $\phi$, 
\postrefereechanges{to what degree should} the optimal cross-correlations 
\postrefereechanges{favour $\phi$ close to} zero, and what shape should the distribution 
of $\phi$
take? For example, if $\sigma_\phi$ as defined above is estimated on
the interval (without periodicity) $[-\pi,\pi)$, then how small should
it be? 

The simulations are generated from spherical harmonics 
using the spherical harmonic mean coefficients $C_l$ of
\nocite{Hinshaw06}{Hinshaw} {et~al.} (2007), obtained from the WMAP data, 
but with random phases 
(and simulated noise).
We choose those of the simulations with the
lowest $\Sxi$ estimates.  These simulations contain much of the
same information that is in the observational data set. After all,
this is the point of simulations.

We know that the MCMC chains in
the data with the correct phases favour $\phi = + \pi/5$, and
have an anti-correlation at $2\rSLS$. Randomising the phases of 
the spherical harmonics while retaining the $C_l$ values should
yield many statistical properties that are similar
to that of the map with the correct phases.
So,  the simulated maps 
must statistically allow $\phi$ to be 
at least as high as $\pi/5$, but do not necessarily have to 
allow it to be much higher. 
Since the random phases have no way of favouring right-handed twists over
left-handed twists, a distribution centred at $\phi=0$ and including values
roughly up to $\pm \pi/5$ is consistent with the information that
the simulations should statistically contain. However, the reason
why the distribution does not extend to optimal twists with
$|\phi| \gg \pi/5$ remains an open question for future work.

\section{Conclusion} \label{s-conclu}

By use of some spherical trigonometry, it is possible to speed up 
the Markov Chain Monte Carlo cross-correlation method of testing a cosmic
topology hypothesis described in RBSG08 by a factor of about 3--10. 
This is shown in
Fig.~\ref{f-alpha_p} and Eq.~(\ref{e-alpha-pm}). 
This could, in principle, make it practical to make 
calculations at higher resolution than before. However, 
the physical interpretation of the calculations would be 
ambiguous because of several systematic effects listed above.

Moreover, for low matched circle sizes and low maximum pair
separations, use of Eq.~(\ref{e-alpha-pm}) can increase 
the numbers of pairs per separation bin, thereby decreasing the noise.
A further improvement in the method and the relation between this
method as a test of ``matched annuli'' and tests of matched circles 
are discussed above in \SSS\ref{s-disc-speed}.

Applying the faster method to the ILC and TOH versions of the WMAP five-year
data, we find little significant change in the best estimate parameters
for a Poincar\'e dodecahedral space model of the Universe, compared to
those given in RBSG08. Depending
on the minimum pseudo-probability level used in the MCMC chains that is
used for estimating the twist $\phi$, the optimal value of $\phi$ 
is a few degrees greater and a few degrees less
than $+36.0\ddeg$ 
in the ILC and TOH maps respectively (Table~\ref{t-alpha-phi-5yr}).

We also applied the faster method to a small number of simulated skies.
The WMAP observations confirmed the generic cosmic topology
prediction \nocite{Star93,Stevens93}({Starobinsky} 1993; {Stevens} {et~al.} 1993) of a cutoff in structure statistics
such as the temperature-temperature fluctuation auto-correlation on large 
angular or spatial scales. Here, we estimated the weakness of the large
scale auto-correlations using a statistic $\Sxi$ [Eq.~(\ref{e-defn-Sxi})] 
similar to that of \nocite{WMAPSpergel}{Spergel} {et~al.} (2003).
The low observed value of $\Sxi$ 
implies that cross-correlations on these scales should usually be weak,
so that MCMC chains as used here should have difficulty finding a region of
parameter space with high optimal cross-correlations. Hence, to conservatively
test the infinite, flat, concordance model hypothesis, on the assumption
that the real observations have a low value of $\Sxi$ due to being a single
realisation of a random process, we analysed simulated skies which use 
the \nocite{Hinshaw06}{Hinshaw} {et~al.} (2007) $C_l$ estimates and randomised phases of the spherical
harmonics. The observational map was calculated from the three least contaminated
frequency WMAP bands, Q, V and W, and the Gaussian random fluctuation 
simulations were created using an equivalent analysis pipeline \nocite{LR08}({Lew} \& {Roukema} 2008).

For simulated and observed skies with the kp2 Galactic cut, these
simulations using the \nocite{Hinshaw06}{Hinshaw} {et~al.} (2007) $C_l$ estimates generally
yield overestimates of $\Sxi$ for the cut sky.  Since we prefer to
analyse the cut sky in order to minimise galactic contamination
effects, it was necessary to select those simulations with the 
lowest values of $\Sxi$, so that the simulations were statistically
as similar as possible to the observations outside of the kp2 cut.

The results of running MCMC chains on the 20 simulations with the lowest
$\Sxi$ estimates from among 50 simulations were that only two
simulations gave optimal cross-correlation solutions which did not
escape to the lower limit in circle size $\alpha$, where small number
statistics favour fortuitous cross-correlations. This indicates a
conditional probability of finding a non-noise solution of about
10\% [Eq.~(\ref{e-prob-PDS-given-Sxi_alpha})].  

The distribution of optimal twists $\phi$ from the simulations
showed that despite the low value of $\Sxi$, the expected distribution
of $\phi$ in the simulations was not uniform. Instead, it is
consistent with a Gaussian distribution centred 
\postrefereechanges{near} $\phi=0$, of
width $\sim 33$--$38\ddeg$. Possible reasons for the nature of this
distribution are discussed above, in
Sections~\ref{s-why-phi-symmetric} and \ref{s-why-phi-distbn}.

Assuming this numerical, Gaussian fit to the expected (for an infinite,
flat model) distribution of $\phi$, and 
using the estimate of $\phi$ obtained
from the observational map (Table~\ref{t-alpha-phi-INC3}) 
that is most discrepant from $\pi/5$, 
we found that $\phi$ could be expected to be as close to $\pm\pi/5$ as the
observational value is close to $+\pi/5$ with a probability of about 22\% 
[Eq.~(\ref{e-prob-PDS-given-Sxi_phi})].

Both of these probabilities are conditional on $\Sxi$ being low, which
itself is unusual for an infinite, flat model 
\nocite{WMAPSpergel,EfstNoProb03b}(e.g.,  {Spergel} {et~al.} 2003; {Efstathiou} 2004).
Hence, for an infinite, flat, cosmic concordance model with Gaussian
random fluctuations, we find that comparing an observational map with
simulational maps gives an estimate of 
the chance of finding both (a) a
large scale autocorrelation as weak as that observed, and (b) a
PDS-like, optimal cross-correlation signal similar to that observed to 
be about 0.015\% or 1.25\%  for the \nocite{WMAPSpergel}{Spergel} {et~al.} (2003) 
or \nocite{EfstNoProb03b}{Efstathiou} (2004) estimates of the probability of low $\Sxi$ 
respectively [Eq.~(\ref{e-prob-final})].

\begin{acknowledgements}

Thank you to Bartosz Lew for numerous helpful and insightful comments,
\postrefereechanges{and to the anonymous referee who provided constructive
and thoughtful recommendations}.
Usage of the Nicolaus Copernicus Astronomical Center (Toru\'n) computer
cluster is gratefully acknowledged.
Use was made of the WMAP data
(\url{http://lambda.gsfc.nasa.gov/product/}), of the
Centre de Donn\'ees astronomiques de Strasbourg 
(\url{http://cdsads.u-strasbg.fr}), of  
GNU Octave command-line, high-level numerical computation software 
(\url{http://www.gnu.org/software/octave}),
the GNU project R environment for statistical computing and graphics
(\url{http://www.r-project.org/}) and the GNU {\sc plotutils} plotting package.

\end{acknowledgements}

\subm{ \clearpage }

\nice{
%

}


\end{document}